\documentclass[final,3p,times]{elsarticle}

\usepackage{amssymb}
\usepackage{multirow}
\usepackage{amsmath}
\usepackage{amssymb}
\usepackage{mathrsfs}
\usepackage{multicol}
\usepackage{caption}
\usepackage{subcaption}
\usepackage{placeins}
\usepackage{algpseudocode}
\usepackage{algorithm}
\usepackage{float}
\journal{Applied Energy}

\begin{document}

\begin{frontmatter}




\title{Shared Learning of Powertrain Control Policies for Vehicle Fleets}

\author[1]{Lindsey Kerbel}
\ead{lsutto2@g.clemson.edu}

\affiliation[1]{organization={Clemson University},
            addressline={4 Research Dr}, 
            city={Greenville},
            postcode={29607}, 
            state={SC},
            country={}}

\author[1]{Beshah Ayalew}
\ead{beshah@clemson.edu}

\author[2]{Andrej Ivanco}
\ead{andrej.ivanco@allisontransmission.com}

\affiliation[2]{organization={Allison Transmission},
            addressline={One Allison Way}, 
            city={Indianapolis},
            postcode={46222}, 
            state={IN},
            country={}}

\begin{abstract}
Emerging data-driven approaches, such as deep reinforcement learning (DRL), aim at on-the-field learning of powertrain control policies that optimize fuel economy and other performance metrics. Indeed, they have shown great potential in this regard for individual vehicles on specific routes/drive cycles.  However, for fleets of vehicles that must service a distribution of routes, DRL approaches struggle with learning stability issues that result in high variances and challenge their practical deployment. In this paper, we present a novel framework for shared learning among a fleet of vehicles through the use of a distilled group policy as the knowledge sharing mechanism for the policy learning computations at each vehicle. We detail the mathematical formulation that makes this possible. Several scenarios are considered to analyze the framework's functionality, performance, and computational scalability with fleet size. Comparisons of the cumulative performance of fleets using our proposed shared learning approach with a baseline of individual learning agents and another state-of-the-art approach with a centralized learner show clear advantages to our approach. For example, we find a fleet average asymptotic improvement of $8.5\%$ in fuel economy compared to the baseline while also improving on the metrics of acceleration error and shifting frequency for fleets serving a distribution of suburban routes. Furthermore, we include demonstrative results that show how the framework reduces variance within a fleet and also how it helps individual agents adapt better to new routes.

\noindent
© 2024. This manuscript version is made available under the CC-BY-NC-ND 4.0 license https://creativecommons.org/licenses/by-nc-nd/4.0/
\end{abstract}



\begin{highlights}
\item A shared learning algorithm for a fleet of vehicles serving a distribution of routes.
\item A characterization of the scalability and computational complexity of the algorithm.
\item Demonstration of the benefits of shared learning for fleets in varying scenarios.
\end{highlights}


\end{frontmatter}

\maketitle

\section{Introduction}
Deep reinforcement learning (DRL) has emerged as a powerful approach to optimize the performance of vehicles in the field. In the context of this paper, DRL broadly refers to a set of approaches that train deep neural networks representing vehicle control policies using experiential data. This approach has immense potential to reduce energy consumption and emissions from individual vehicles by exploiting the granular data that can be logged from vehicles via connectivity and/or edge devices. The relevant data that can be used optimizing and adapting powertrain control (PTC) with DRL include fuel rate, engine speed and torque, gear position, driver pedal positions, and so on. However, successfully training and deploying DRL frameworks on real vehicles has challenges of data efficiency and learning stability, while also maintaining safety. These challenges emanate from the needs of DRL approaches to do extensive and diverse exploration to learn good policies. In this paper, we seek to address these challenges by leveraging the experiences of a fleet of vehicles to learn optimal PTC policies. We approach this by proposing a shared learning framework that allows vehicles to exchange information through a shared group policy.

For the purposes of this paper, a PTC policy means the composite decision made at each time step for discrete transmission shifting (gear change) and continuous IC engine traction/brake torque control. However, other types of powertrain configurations such as pure electric and hybrid vehicles with fewer or more decision variables can also be readily accommodated with the approach presented. We define the term 'fleet' as a group of vehicles, potentially owned by an enterprise, with each vehicle assigned to independently operate on a given route for a work shift in response to potentially random customer demands. For instance, this scenario applies to a logistics company delivering goods to various customers on a given distribution of routes (by speed/traffic profile, road topology, loading, etc.).

In recent times, DRL has been successfully demonstrated on various vehicular applications. For instance, DRL techniques have been proposed for energy management systems for hybrid electric vehicles (HEV)~\citep{Hems2}. DRL has also been proposed for various autonomous vehicle maneuvers, such as car following scenarios~\citep{AVfollow}. Another notable application includes gearshift control for automatic transmissions~\citep{transmission}. In our own initial application of DRL for PTC (joint gear shifting and torque control) of a commercial vehicle, we observed fuel savings of up to $12\%$ over established table-based approaches~\citep{kerbel}, which is comparable to the findings of a study that utilized dynamic programming where the vehicle model and environment were assumed deterministic and known in advance~\citep{dp_pt}. Although many of these works have demonstrated successful learning of vehicle control policies (many in simulated environments), they have shown limited adaptability to diverse real-world cycles~\citep{adapt}. The challenges arise from the highly dynamic and complex nature of a vehicle's environment, encompassing factors such as traffic, unknown routes, and vehicle dynamics. The necessary training time to gather diverse data and the computational cost of learning an implementable DRL policy that covers this complexity is prohibitively high~\citep{traffic_challenges}.

To mitigate some of these inherent challenges of DRL training for real-world scenarios, various approaches have been proposed. Many methods seek to utilize external knowledge to reduce training time, reduce unnecessary or unsafe exploration and/or enhance learning stability. For example, applying imitation learning using human driving data has been shown to improve stability of the learned policy~\citep{auto_imitation}. Human-in-the-loop guidance has also been proposed as a real-time solution to improving autonomous driving policies~\citep{AV_Human}. Another approach couples DRL with model predictive control for HEV energy management~\citep{Hems_MPC}. Our own approach is broadly related to a group of methods called transfer learning, which typically seek to leverage or adapt existing knowledge of source or expert policies to the situation at hand. Therein, some approaches learn attention/weight networks to selectively use actions from the source policies and add them to the policy actions being learned~\citep{a2t}. Similarly,~\citep{multipolar} applies an additional learned residual policy with an attention network. Along these lines, we had proposed leveraging the default PTC policy (shipped with the vehicles) to enable learning on the field via residual policy learning~\citep{kerbelRPL}. We then proposed an attention network to speed up learning and offer adaptability to changes in the environment (vehicle mass, driver, routes, etc) for individual vehicles~\citep{APLkerbel}. Other transfer learning approaches, which are broadly called policy distillation approaches~\citep{distillation}, seek distribution matching between the target policy and one or more source (expert) policies~\citep{distilling}. Furthermore,~\citep{dualdistillation} aims to match distributions between two learning agents rather than a source policy. In this paper, we draw from these policy distillation notions to leverage the collective experience of vehicles in a fleet to improve the learning performance of all vehicles on the distribution of routes serviced by the fleet. The resulting framework does not use explicit source policies.

Our approach will also draw from developments in \textit{multi-task RL} (MTRL), which encompasses strategies that enable a DRL agent to learn a policy capable of solving multiple tasks effectively. One can consider categories of driving cycles (urban, highway, mostly flat, mostly hilly, and any combinations thereof) as different tasks. In the broader literature, chiefly in robotics, MTRL entails training on different specific tasks in parallel and then combining the task-specific policies into a generalized multi-task policy~\citep{1997multitask}. One such method, known as Importance Weighted Actor-Learner Architecture (IMPALA), uses a centralized actor-critic that learns from the combination of experiences in individual tasks~\citep{impala}.~\cite{actormimic} uses supervised regression from multiple expert policies to learn a generalized multi-task policy. Distill and Transfer Learning (Distral)~\citep{distral} extends this by then using the multi-task policy to regularize the task-specific learners towards the shared multi-task policy.~\cite{divide} propose a divide and conquer approach where they constrain the learning of the task-specific policies to each other and then regularly reset the task policies to a learned multi-task policy. We adopt the notions of a learned shared policy from these works on MTRL to the present vehicle PTC application. However, instead of seeking to make each vehicle's PTC policy a multi-task policy in the sense used therein, we make each vehicle's PTC agent learn its own local policy using the experiences it generates itself. Then, from these local policies of the vehicles in the fleet, we regress a shared group policy, towards which the individual agent policies are subsequently regularized. This will have the desired effect of learning a multi-task policy by the individual vehicles as we will demonstrate with our experiments. In this vein, our method also has similar objectives as meta-learning methods~\citep{maml}.~\cite{metalearning} seeks domain generalization, which is similar in terms of vehicle agents learning a robust and adaptable policy on a distribution of tasks/routes.

Our interest on learning via a fleet of vehicles also invokes \textit{Multi-agent RL} (MARL) techniques. However, the most common formulations of MARL are about agents that act and learn in a shared physical environment and usually involving a shared reward/objective. This is not quite the case for a fleet of road transport vehicles considered in this work, where the agents are generating experiences on their individual routes. The vehicles in the fleet are not expected to have spatial or temporal proximity during their exploration and learning. Nevertheless, some of the MARL notions of centralized training with decentralized execution, such as the multi-agent Q-learning proposed in~\cite{qmix} and actor-critic architectures with a common critic and independent actors~\citep{coma} could potentially be adopted to the present application. Indeed such MARL has been applied to improve fleet dispatching and routing~\citep{marl_routing}. Furthermore, it has been applied for energy scheduling and routing of electric vehicle (EV) fleets~\citep{marl_schedule}. MARL is also commonly employed to coordinate cooperative maneuvers among autonomous vehicles, such as velocity control in multi-vehicle scenarios~\citep{marl_auto}. The present work seeks to retain independent learning by vehicle agents (using any suitable local DRL algorithm) while sharing knowledge via a centralized 'fleet coordinator' that computes a shared group policy.

Our shared learning framework also has similarities to federated reinforcement learning (FRL), which is popular for connected autonomous vehicle applications that require shared sensor data to remain private~\citep{FRL_auto}. These approaches typically involve aggregating network weights or sharing of gradient information, ensuring that sequential data is not directly shared~\citep{FedAvg}. In our proposed approach, information is exclusively shared with the fleet coordinator, and each vehicle only receives the (parameters of the) group policy network (actor), and shares its own policy and value network parameters, and randomized state information from its experience buffer. No sequential state information is shared that can readily compromise privacy. Therefore, although data privacy is not a focus of this work, our proposed framework has some safeguards advocated in FRL works.

In this paper, we develop a scheme for learning a shared group policy from the individual vehicular agents' localized policies and subsequently use the group policy to constrain the learning at the individual agents. The group policy ($\pi_g$), which can be hosted on a centralized fleet coordinator, is configured to capture the common aspects of the policies learned by the individual vehicle's PTC agents, which are the learner-explorers in our setting. The group policy is only used to constrain training of the individual agent policies ($\pi_i$) towards common behaviors (action choices) in similar states, and therefore does not interact with the environment. We compare our approach with a modification of the state-of-the art IMPALA algorithm~\citep{impala} that uses a centralized actor-critic on shared data, as well as with the baseline of individually learning PTC agents without any sharing. 

The main contributions of this paper include: 
\begin{itemize}
\item A mathematical framework and a detailed algorithm for shared policy learning by a fleet of vehicles assigned to a distribution of routes.
\item Characterizations of the computational requirements and scalability of the approach considering computation time and data traffic.
\item Demonstration of the performance of the framework on simulations of a fleet of commercial vehicles in multiple routing scenarios  including comparisons with the no-sharing baseline as well as with a state-of-the-art algorithm that learns  a centralized policy from shared data.
\end{itemize}

The rest of this paper is organized as follows. Section \ref{sec:problem} presents the problem formulation at a high level. In Section \ref{sec:algo}, we present the mathematical details of the proposed shared learning framework and corresponding algorithms. Section \ref{sec:experiment} details the experimental setup for the simulation and discusses the scenarios presented.  The results for three scenarios and discussion of the benefits of shared fleet learning are presented in Section \ref{sec:results}. That section also discusses the scalability aspects and computational costs of the algorithm. Section \ref{sec:conclusion} summarizes the main points and concludes the paper.

\section{Problem formulation} \label{sec:problem}
\subsection{Independent learning on random routes}
Consider PTC for a fleet of vehicles that are assigned different routes in response to customer pickup and delivery or service orders. The latter are often random, therefore each vehicle can be considered to take routes in a correspondingly random fashion (say, per work shift) by sampling from the available delivery/pick-up routes. To simulate this, we say each vehicle samples and learns from a distribution of routes $R_i\sim{}p(R)$. These routes may be characterized by speed profiles, grades, traffic patterns, loading, etc. Each vehicle trains a DRL agent for its PTC on the routes it samples. Given this practical scenario, it is desirable to achieve \textit{flexible utilization of the fleet} of vehicles and robustly respond to stochastic customer requests with minimal variability in performance. This can be viewed as training the fleet of vehicles for the distribution of tasks with multi-task policies. 

\begin{figure}[h]
      \centering{
	\includegraphics[width=\textwidth]{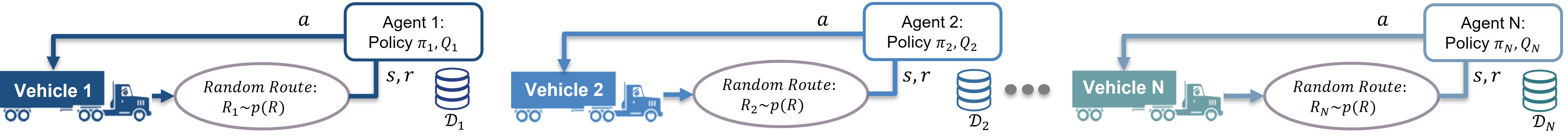}}
      \caption{Fleet of vehicle DRL agents independently learning}
      \label{fig:formulation}
 \end{figure}

We start with the baseline scenario, depicted in Figure \ref{fig:formulation}. Therein, the DRL agents for each of the vehicles in the fleet learn/train independently on the routes they sample. This is an extension of the commonly used approach of training on a single route to optimize route specific performance. The distinction here is that when routes are sampled randomly as depicted in Figure \ref{fig:formulation}, the learning performance is severely degraded as we will demonstrate in Section \ref{sec:results}. The definitions of the Markov Decision Processes (MDPs) for individual vehicle DRL agents have been covered extensively in many prior works~\citep{ptcmdp}. In Section \ref{sec:experiment} and Appendix A, we offer a brief discussion of the individual vehicle DRL PTC agent algorithm and the selected training scheme we use for the results in this paper. However, we note that many other state-of-the-art DRL algorithms can be used at the individual vehicle level and fit into our shared learning framework. We give further insights on this in Section \ref{sec:rl_agent} as well. 

\subsection{Shared policy learning framework}
\label{sec:framework}
\begin{figure}[h]
      \centering{
	\includegraphics[width=.65\textwidth]{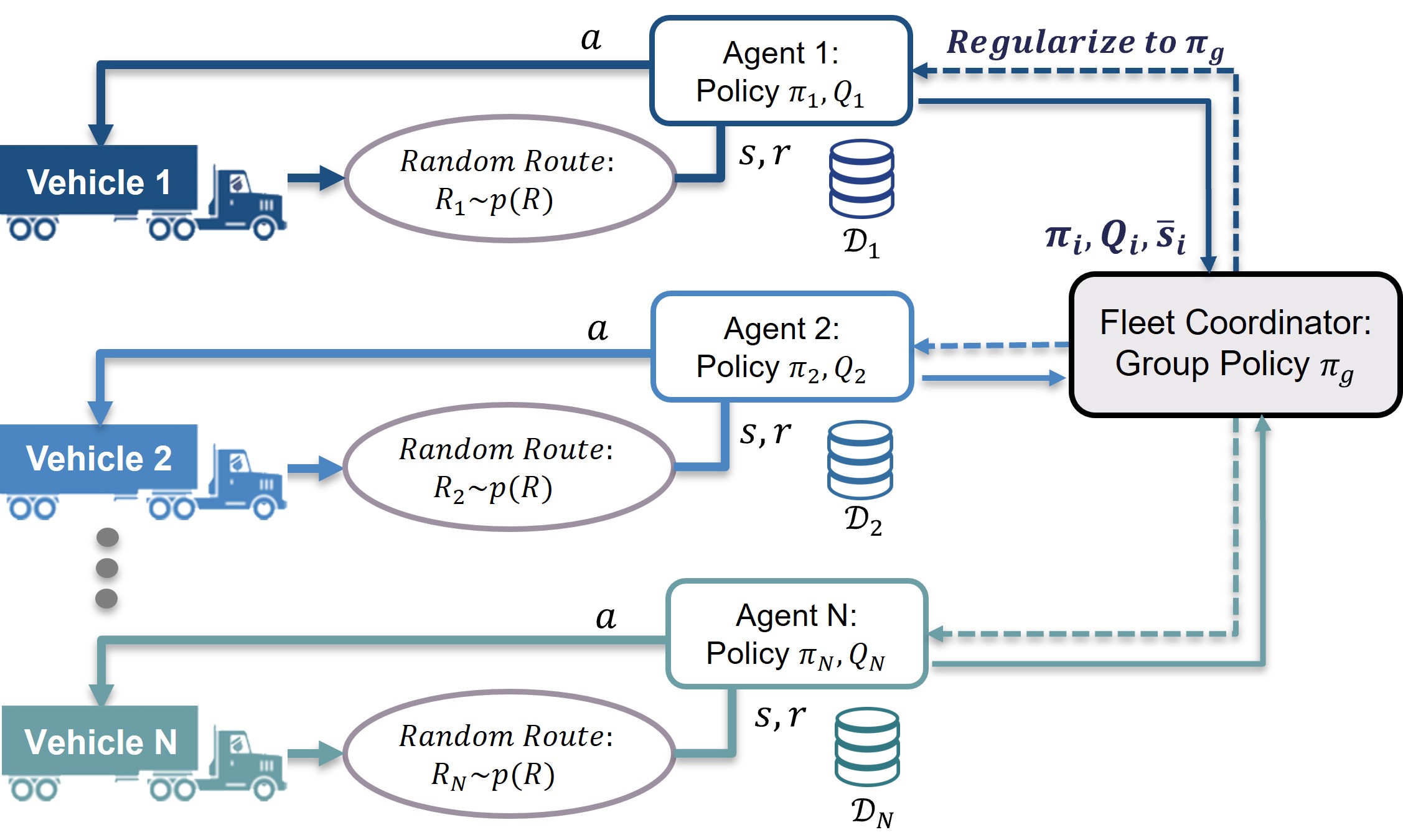}}
      \caption{Framework for sharing learning using a group policy}
      \label{fig:central_framework}
 \end{figure}

Our main proposed approach for shared fleet learning is illustrated in Figure \ref{fig:central_framework}. The key addition to the baseline scenario of independently learning agents is the creation and learning of a group policy ($\pi_g$) as the coordination mechanism. As the group policy is updated, it is sent to the individual vehicles which use it to regularize (constrain) training of the individual agent policy ($\pi_i$). This process can be executed \textit{asynchronously} to use the latest policy updates both at the group and agent levels. It ensures that agents can share their latest policies without any restrictions on when information is exchanged. However, to simplify discussions and notations, we assume synchronous updates of both the group and agent policies at regular intervals in the discussions below.

We opted to utilize policy information instead of the state-action values (Q-values estimated by DRL agent critics for the individual vehicles) as the main mechanism for knowledge transfer. This is because Q-values are influenced by scaling of the agents' rewards which may vary significantly between vehicles (physics, drivers, etc) and routes (tasks) and their estimates often exhibit high variance. The policies, on the other hand, are action probabilities (for torque and gear selection for the examples in this paper) that can be readily regressed over to learn a group policy. However, shared learning is made more effective by weighing individual policies via their estimated advantage functions, which are derived from Q-values, as we describe below. Therefore, Figure \ref{fig:central_framework} shows that the individual agent Q-values, $Q_i$, the agent policy, $\pi_i$, and randomly sampled states, $\bar{s}_i$, visited by agent $i$, are sent to the fleet coordinator which computes the group policy. Only the group policy, $\pi_g$, is then shared back to each indvidual agent.

\section{Mathematical details of shared policy learning}\label{sec:algo}
The overall objective function to be maximized for the group (fleet) of agents participating in the shared learning scheme can be written as:
\begin{equation} \label{eqn:overall}
\mathcal{J}_{\pi_g, {\pi_i}_{i=1}^N} = \sum_{i=1}^N \mathbb{E}_{\pi_i} \left[ \sum_{t=0}^{\infty} \gamma^t r_i(a_t, s_t) + \gamma^t \zeta_i \log{\pi_g(a|s_t)} - \gamma^t \lambda_i \log{\pi_i(a|s_t)}\right]
\end{equation}
where the group policy is designated by $\pi_g$ and each agent's policy by $\pi_i$ (See Figure \ref{fig:central_framework}). The first term is the standard discounted return following each agent's policy, the second term can be considered as a reward shaping term encouraging actions of high probability under the group policy, and the last term is an entropy regularization term to encourage exploration by the agent policies. $\zeta_i$ and $\lambda_i$ are weight factors on the contributions of the corresponding terms. As in standard DRL, we assume that future rewards are discounted by $\gamma$. The above objective is inspired by the MTRL proposal of Distral~\citep{distral}, with the key difference that we do not consider individual vehicle agent’s policies as task-specific policies in the sense used there.

While it is possible to optimize the above objective function jointly for the agent and group policies, we take an alternating approach that allows an asynchronous implementation of the agent and group policy learning using the latest available iterates for each. That is, we consider the optimization of the objective in (\ref{eqn:overall}) by alternating optimization with respect to $\pi_i$ and $\pi_g$ (fixing one, then the other). In particular, the optimization with respect to the group policy $\pi_g$ becomes a supervised regression problem, namely one of maximizing the log likelihood of a model for $\pi_g$ for state visitations under the agent policies $\pi_i$ in the routes $R_i$ that the agent samples. Furthermore, in our implementation, the individual vehicle's agents (not the group policy) act in the environment and the evaluation of their performance is estimated through their individual critics. We use this information to further enhance the learning of the group policy $\pi_g$ via an advantage weighted regression that we detail below. This is another distinction versus the proposal in~\cite{distral}.

\subsection{Learning individual agent policies in the shared setting}\label{sec:rl_agent}
While many modern DRL setups and algorithms can be configured with the proposed shared policy learning scheme, we build on an actor-critic DRL architecture developed in our prior work for individual vehicle PTC. We briefly summarize the key attributes here and defer a more detailed discussion to prior work~\citep{APLkerbel} and the Appendix.

The individual vehicle PTC agent is constructed as an eco-driver assist DRL agent. We train the DRL PTC agent by considering its interaction with the environment (consisting of the driver, vehicle and traffic) as an MDP. Specifically, the agent uses a scalarized multi-objective reward function penalizing acceleration error versus driver demand, fuel rate, gear shifting frequency and power reserve terms. The specific form of the reward function used for our experiments is given in Section \ref{agentsetup}. The PTC actions/decisions made by the DRL agent are the discrete gear change action $a^d$ and the continuous traction torque $a^c$ for a commercial vehicle with an automated transmission and IC engine powertrain. Considering that these two actions are independent, the DRL PTC policy $\pi_i$ can be factorized as $\pi_i(a|s) = \pi_i^c(a^c|s)\pi_i^d(a^d|s)$, where $\pi_i^c$ and $\pi_i^d$ are the continuous and discrete policies, respectively. We use a deep neural network to represent the policy $\pi_i$, where we model the discrete gear policy $\pi_i^d$ as a categorical distribution via a softmax output layer and the continuous torque policy $\pi_i^c$ as a Gaussian distribution with mean $\mu_i$ and standard deviation $\sigma_i$. 

To train the individual vehicle DRL agent policies $\pi_i$, we adopted the Maximum A Posteriori Policy Optimization (MPO) algorithm~\citep{abdolmaleki2018}. In our experiments, it offered robustness and minimal hyperparmeter tuning versus other state-of-the art algorithms which are also applicable in principle, such as Deep Deterministic Policy Gradient (DDPG)~\citep{ddpg} for example. We opted for an off-policy version to allow training on asynchronous data from prior policy iterates. Furthermore, MPO allowed a straight-forward implementation for the hybrid-action space of the present application~\citep{neunert2020}. We give a brief explanation of the MPO algorithm in the Appendix. We apply MPO to train the DRL agent policy for each individual vehicle and we denote the algorithm's loss function (optimization objective for individual vehicle DRL without any sharing) as $\mathcal{L}_{base, i}$. We remark here that our implementation of MPO does use a critic network to also learn the state-action values for each vehicle agent, denoted by $Q_i$. These will be used for the shared learning as already mentioned above and described further in Section \ref{group_learning} below.


Returning to the shared learning framework of the present paper, the goal is to regularize each agent's policy towards the group policy to maintain common behaviors. To achieve this, we modify the base loss function $\mathcal{L}_{base, i}$ of the individual vehicle DRL agent by adding a Kullback–Leibler (KL) divergence regularization which penalizes the deviation of the agent’s policy $\pi_i$ from the group policy $\pi_g$ and ensures that the agents’ policies align more closely with the group policy. The new loss function for each individual vehicle DRL agent is  given by:
\begin{equation} \label{eqn:agent}
\begin{split}
\mathcal{L}_{i} &= \mathcal{L}_{base, i} + \lambda_i  KL\left(\pi_g\left(a|s\right)||\pi_i\left(a|s\right)\right)\\
&= \mathcal{L}_{base, i} + \pi^d_g(a^d|s) \lambda_{c,i} KL\left(\pi^c_g\left(a^c|s\right)||\pi^c_i\left(a^c|s\right)\right) + \pi^c_g(a^c|s) \lambda_{d,i} KL\left(\pi^d_g\left(a^d|s\right)||\pi^d_i\left(a^d|s\right)\right)
\end{split}
\end{equation}

Note that the loss function in Eq. (\ref{eqn:agent}) is optimized with respect to the agent policy components $\pi_i^c$ and $\pi_i^d$, while the group policy components are held fixed. In the MPO algorithm for optimizing Eq. (\ref{eqn:agent}), the individual KL divergences are enforced as separate terms for additional control over the regularization of the respective components. We remark that we opted to use the moment-matching forward KL divergence to consider the full support of the group policy distribution~\citep{forwardkl}. This is in contrast to other works that use the reverse KL (mode-seeking) for policy distillation, including Distral~\citep{distral}. Reverse KL tends to put probability mass on the highest probability actions which tends to limit exploration and lead to sub-optimal solutions. In most cases, however, either of these divergences give similar results. 

Next we discuss how to obtain the group policy. 

\subsection{Advantage weighted regression to learn a group policy}\label{group_learning}
As described above, to compute or learn the group policy $\pi_g$, we minimize the cross-entropy loss function given by Eq. (\ref{eqn:group}). This is simply a supervised regression over the current iterates of the individual agent policies.
\begin{equation} \label{eqn:group}
\mathcal{L}_{\pi_g} = -\sum_{i=1}^N \mathbb{E}_{\pi_i(a|s)}\log{\pi_g(a|s)}
\end{equation}
For our discrete-continuous hybrid action space for PTC, we pursue the optimization by resolving the cross-entropy loss in Eq. (\ref{eqn:group}) into corresponding components in a similar vein as above. The cross-entropy loss function over all the discrete (gear) action choices is given by:
\begin{equation} \label{eqn:reg_disc}
\mathcal{L}_{\pi_g,d_i} =  -\sum_{a^d \in \mathcal{A}^d} \left[\pi_i (a^d|s) \log{\pi_g (a^d|s)}\right]
\end{equation}
where $\pi_i$ and $\pi_g$ are the probabilities given by categorical distributions (softmax output) of the current iterates of each individual agent policy, and the group policy, respectively. To compute the cross-entropy for discrete actions, we consider the probabilities of all possible action choices, denoted by $a^d \in \mathcal{A}^d$. While an alternative option is to use the one-hot vector representing the policy action chosen, we observed empirically that this approach caused the group policy distribution to heavily favor a single action during early training. Consequently, it led to a similar shift in the local agent policies, limiting exploration. 

For the continuous torque action, we first recall that the agent policies $\pi_i$ are parameterized as Gaussians with mean ($\mu_i$) and standard deviation ($\sigma_i$). We use the following closed form cross-entropy loss for Gaussian distributions to learn a mean ($\mu_g$) and standard deviation ($\sigma_g$) of the group policy.
\begin{equation} \label{eqn:reg_cont}
\mathcal{L}_{\pi_g,c_i} = \frac{1}{2}\log{2\pi\sigma_g^2} + \frac{\sigma_i^2 + \left(\mu_i - \mu_g\right)^2}{2\sigma_g^2}
\end{equation}
Simply combining the loss functions in Eq. (\ref{eqn:reg_disc}) and Eq. (\ref{eqn:reg_cont}) would weigh the contributions of individual agent policies equally. However, because of the stochastic nature of distributional policies, we cannot always expect all agent policies to be consistently good. This is further exacerbated in the fleet setting as we consider vehicles/dispatchers to randomly sample routes to respond to random customer demands. As each agent’s policy may not be optimal at every iteration in this setting, we need a way to assess the quality of the actions chosen by individual agents so we can weigh their contributions to the loss accordingly. The advantage function, expressed by $A_i(s, a) = Q_i(s, a) - V_i(s)$, is commonly used in DRL to favor actions that outperform the average action in a given state~\citep{qweighted}. Herein, $Q_i(s, a)$ is the state-action value, as estimated by the critic network, and $V_i(s)$ is the state value. While the Q-values are output from the individual DRL agents (see Figure \ref{fig:central_framework}), the state values $V_i(s)$ are estimated by averaging over the agent's Q-values for a sampled set of all the available actions in that state. This is done for the batch of states sampled from the agent's memory buffer (common in DRL implementations) and sent to the fleet coordinator.
\begin{equation}\label{eqn:value}
V_i(s) =  \frac{1}{M} \sum_{j=1}^M Q_{i}(s,a_j)
\end{equation}
where $M$ is the number of actions sampled from the entire action space for each sampled state ($s$). To quantify the advantage of the current agent's policy actions compared to the value of all the possible actions in each state, we use Q-values for the greedy (maximum probability) actions ($a_g$) from the agent's current policy. We use an exponentiated version of the advantage function to overcome varying scale issues with value functions~\citep{AWR}. The final advantage-weighting factor is given by: 
\begin{equation}\label{eqn:advantage}
\zeta_i = \exp{\left(\frac{A_i(s,a_g)}{\beta}\right)} = \exp{\left(\frac{\left[ Q_i(s,a_g) - V_i(s)\right]}{\beta}\right)}
\end{equation}
where $\beta$ is a temperature hyperparameter that can be tuned for all agents.

Finally, we sum each agent's discrete and continuous loss functions weighed by the exponentiated advantage function to obtain the loss function for the group policy ($\mathcal{L}_{\pi_g}$):
\begin{equation} \label{eqn:regression}
\mathcal{L}_{\pi_g} = \sum_{i=1}^N{\zeta_i\left[\mathcal{L}_{\pi_g,d_i} +\mathcal{L}_{\pi_g,c_i}\right]}
\end{equation}
We used the stochastic gradient-based optimization solver Adam~\citep{Kingma2015AdamAM} to optimize the loss functions (Eq. \ref{eqn:agent}, \ref{eqn:regression}) given above with respect to the neural network parameters for the group policy, as well as the individual agent policies and their critics (via the MPO algorithm).

\subsection{Shared Policy Learning Algorithm Pseudocode}
The pseudocode below outlines the steps for updating the individual agents' policies (left) and the group policy (right) for the shared learning framework. The algorithms can be executed asynchronously using the most recently updated policies. The trajectory of experiences for each agent $T_i$ is saved in the corresponding agent's memory buffer $D_i$. The number of batches sampled is $n$ and the size of the batch is $B$ experiences (states, actions, rewards, and next states). Note that the individual agent pseudocode can be implemented with many modern DRL algorithms. 
\begin{algorithm}
\caption{Shared policy learning pseudocode}\label{alg:cap}
\begin{multicols}{2}
\begin{algorithmic}
\Ensure $Q_i, \pi_i = \pi_g$
\State \textbf{Individual agent learning (at each agent $i$):}
\State Collect trajectory $T_i$ by applying $a_i$ from $\pi_i(a|s)$ and store in $D_i$
\State Sample $n$ random batches of size $B$ from $D_i$
\For{each batch}
	\State Update the critic $Q_i$
	\State Solve for agent loss $\mathcal{L}_{\pi_i}$  using recent  $\pi_g$ (Eq. $\ref{eqn:agent}$)
	\State Update $\pi_i$ with gradient step from $\mathcal{L}_{\pi_i}$ 
\EndFor
\State Share $\pi_i$, $Q_i$, sampled batches $\bar{s}_i$ to fleet coordinator
\end{algorithmic}
    \columnbreak
\begin{algorithmic}
\State \textbf{Group policy learning:}
\For{all agents $i$}
\State Get recent policy $\pi_i$, $Q_i$, and sampled batches $\bar{s}_i$
\State Solve for advantage weight $\zeta_i$ with recent $Q_i$
\State Solve for $\mathcal{L}_{\pi_g,d_i}$ and $\mathcal{L}_{\pi_g,c_i}$ with recent $\pi_i$ (Eq. $\ref{eqn:reg_disc}, \ref{eqn:reg_cont}$)
\EndFor
\State Solve for $\mathcal{L}_{\pi_g}$  (Eq. $\ref{eqn:regression}$)
\State Update $\pi_g$ with gradient step from $\mathcal{L}_{\pi_g}$ 
\State Share $\pi_g$ to all agents
    \end{algorithmic}
  \end{multicols}
\end{algorithm}

Although using the most recent experiences of the individual vehicle agents seems intuitive for the group policy regression, we empirically found that sampling a randomized batch of experiences from each agent’s memory buffer leads to more stable learning than using only the most recent experiences to fit the group policy (on the right). Such random sampling is known to mitigate the bias of learning only from the most recent experiences. Additionally, this approach addresses data privacy concerns by sharing non-sequential state information, as pointed out earlier. 

\section{Experimental Setup}\label{sec:experiment}
We demonstrate and evaluate our shared learning framework through simulations of fleets of commercial vehicles implementing DRL for their PTC. All simulations are conducted on the Palmetto Cluster~\citep{cluster} using Python scripts, and the DRL agents are built using tools in PyTorch. To evaluate computational timing aspects we use a dedicated GPU described in Section \ref{sec:results}. In this section, we first detail the simulation setup, providing a brief discussion of the vehicle and driver models, followed by the configuration of individual DRL agents. Then we outline three distinct scenarios we use to demonstrate the workings and performance of the proposed shared learning framework.

\subsection{Environment model for simulation}\label{envsetup}
Consider a fleet consisting of $N$ commercial vehicles with a conventional internal combustion engine and 10-speed automated transmission. We model the vehicles for simulation with the following longitudinal dynamics equation:  
\begin{equation} \label{eqn:vehicle}
\frac{dV_e}{dt}=\frac{1}{M_{eff}}\left[\frac{T_t}{r_w} +R_r (S_e)+R_a (V_e)+R_g (S_e)\right] 
\end{equation}
where $R_r(S_e)=WC_{r} \cos \psi (S_e)$, $R_a (V_e)=0.5 \rho C_d A_f V_e^2$, and $R_g (S_e)= W_e \sin \psi (S_e)$. $A_f,C_{1}$, $\rho$, and $\psi$ are the frontal area of the vehicle, the coefficient of rolling resistance, the air density, and the road grade as a function of the vehicle's position, $S_e$, respectively. The vehicle’s velocity, weight, effective mass, and wheel radius, are listed as $V_e,W_e,M_{eff}$ and $r_w$, respectively. The traction torque $T_t$ is the sum of any positive and negative torques applied to the wheels from the engine and/or the service brakes.  A negative traction torque is distributed first by applying the maximum engine braking allowed and then the remaining torque is applied by the service brake system at the wheels. The fuel rate is interpolated from a table given as a function of engine speed and torque. 

To simulate the driver, we use the Intelligent Driver Model (IDM), a well-established model for simulating typical car-following behavior~\citep{TreIDM}. The IDM outputs the driver's desired acceleration for the ego-vehicle as a function of the relative distance and relative velocity of the ego-vehicle with respect to the preceding vehicle. This model's parameters can be varied to simulate variations in driver tendencies. The leading vehicle follows a given velocity profile, which we refer to as the route. 

\subsection{Individual vehicle DRL agent setup}\label{agentsetup}
We set up an MDP for each individual vehicle's DRL PTC agent. The state vector is formulated as $s_t=[V_{e, t}, A_{e, t}, A_{des, t}, n_{g,t}, A_{des, t-1}, n_{g, t-1}]$, where $V_e$ is the ego vehicle's velocity, $A_e$ is the ego vehicle's acceleration, $A_{des}$ is the driver's desired acceleration (modeled via IDM, but in practice expressed through the brake and acceleration pedals), and $n_g$ is the transmission gear. The subscript $t$ refers to the values at the current time and $t-1$ are the values from the previous time step. We note that the DRL PTC agent in this study does not use the leading vehicle information directly. It instead uses the driver's desired acceleration ($A_{des}$) as one of the elements of the state information. The action output vector includes a torque and gear selection formulated as $a=[T_{t}, u_g]$, where $T_{t}$ is the desired wheel torque and $u_g$ is the gear change command with choices setup as $[-1,0,1]$ (downshift, remain in current gear, or up-shift). The vehicle system (in the environment) enforces operable actions consistent with the limitations of the powertrain and applies these actions to the vehicle model to transition to the next state, generating the rewards as defined below. Since individual vehicle agents operate independently, their experiences are stored in separate local memory buffers. 

To define the reward function for the MDP, we consider the DRL PTC agent to be a driver-assist controller (that takes in the driver demand to act on the powertrain). The primary goal is to balance energy consumption, driver demand accommodation, and maintain a reasonable level of control activity without compromising on driver comfort. The reward function that encapsulates this goal is a scalarized multi-objective function as expressed in Eq. (\ref{eqn:rewards}). The DRL agent's task is to follow the driver's desired input by minimizing the absolute error between the desired acceleration and the achieved acceleration $(A_{des}- A_e)$, while also minimizing the fuel rate, the gear shifting frequency and traction torque. Another aspect of driveability is to ensure sufficient engine acceleration capability available to the driver, achieved through the consideration of a power reserve ($P_r$) term. To simplify interpretations of weight selections, the components of the rewards are normalized based on their corresponding maximum values.
\begin{equation} \label{eqn:rewards}
\begin{split}
r(s_t, a_t) =&  -W_A\frac{|A_{des_t} - A_{e_{t+1}}|}{\Delta A_{max}} - 
W_{fr}\frac{\dot{m}_{f_{t+1}}}{\dot{m}_{f,max}} - W_g|n_{g_{t+1}} - n_{g_t}| - W_{Tt}\frac{|T_{t_t}|}{T_{t,max}}
- W_{pr}\frac{P_{r, max_{t+1}} - P_{r_{t+1}}}{P_{r, max}}
\end{split}
\end{equation}
where $\Delta A_{max}$, $\dot{m}_{f,max}$, ${T_{t,max}}$, and $P_{r, max}$ are the maximum acceleration error, fuel rate, allowed control torque, and power reserve $P_{r, max}$ as a function of velocity and maximum engine torque. We make each objective a negative reward to discourage (minimize) each of the components as the objective for a DRL agent is often posed as one of maximizing total (discounted) rewards. The weights are tuned once for one representative agent and kept constant across all agents to keep the focus on the evaluation of the shared learning framework.

Since safety is one of the critical concerns limiting DRL for real-world applications, we give a brief remark here. Our DRL formulation is for driver-assistance in which the driver acts to generate the desired acceleration to maintain a safe distance to the leading vehicle according to the IDM (using relative velocity and relative distance). The DRL agent only has access to the desired acceleration and is subordinate to the driver. However, while some acceleration error is unavoidable due to the vehicle dynamics and system constraints (powertrain capabilities), large acceleration error magnitudes that can arise from poor performance of the DRL agent in tracking the driver's desired acceleration are unacceptable. There are indeed explicit steps that can be taken to include safety considerations via control barrier functions (CBF)~\cite{safe_RL2}. Another approach has a distinct critic to evaluate safety constraints~\cite{safe_critic}, although additional relative velocity/distance signals (radar) would need to be assumed available to the safe RL agent (not just the driver). In our own prior work~\cite{safe_RL1}, we demonstrate that safety filters based on CBF provide a better safety guarantee than adding reward shaping terms to Eq. (\ref{eqn:rewards}) for safety. The latter merely requires dedicated training without guarantees. For the purpose of our study, we note that any DRL algorithm (with or without explicit/implicit safety terms) can be used by the local agents participating in our shared learning framework. Still, even for the MDP formulation with the reward in Eq. (\ref{eqn:rewards}), we later show in the results that the proposed shared learning framework serves to reduce explorations by limiting the acceleration error variances, thereby helping with safety in that regard, although it doesn’t guarantee it. The combination of safety filtering and shared learning can be explored further in a future work.

The DRL PTC agent for each vehicle in the fleet trains its own actor and critic neural networks, each consisting of 3 linear hidden layers with 256 units per layer. All agents' policy networks, including the group policy model, are initialized with an identical suboptimal policy, rather than starting from scratch. The reasoning for this is to simulate the scenario of agents starting with a pre-existing group policy, reminiscent of an OEM PTC policy shipped with the vehicles, that will then be refined with the actual experiences of the vehicles on the field. Each training route is $10000$ seconds with all agent network updates (learning updates) occurring every $2500$ seconds. The vehicle model parameters and DRL hyperparameters used in this study are listed in Appendix B. 

\subsection{Routing scenarios for vehicle fleets}\label{sec:routingscenarios}
Using a set or distribution of routes, each representing a different velocity profile, we model a multi-task problem that illustrates the difficulty of applying DRL in high variance environments. To demonstrate both the advantages and challenges of the shared learning framework, we simulate three distinct routing scenarios. In this context, a fleet denotes a group of vehicles undergoing training and learning on a specified set of routes ($R$) for the corresponding scenario. During a training cycle, each vehicle in the fleet is arbitrarily assigned one of the routes, resembling a delivery or pick up work shift. Technically, the preceding vehicle is assigned the route, and the ego-vehicle follows the desired acceleration from the IDM, as previously discussed. To introduce additional variability during training, we randomize the velocity profile, driver behavior model, and vehicle mass for each route by incorporating noise into their respective parameters. While road grade is omitted in this work, it could be easily considered as an additional aspect for causing variations in the set of routes ($R$).

In all scenarios, we also conduct comparisons with an alternative shared learning method, employing a modified version of IMPALA~\citep{impala}, as mentioned in the introduction. IMPALA uses a centralized learner with an actor-critic architecture, learning directly from the data/experiences of all the vehicles (agents) which act in their own environments using the central learner's policy. We apply these principles of IMPALA using the same Q-retrace critic and MPO algorithm for its central actor-critic learner as our shared learning agents. After each training route is completed, the agents' policies are reset to align with the updated group policy. In addition to this implementation of IMPALA, we consider a modified implementation of IMPALA where we introduce a modification that each vehicle agent updates their version of the central policy on its training route using its own local actor-critic agent. This modification allows for localized learning by each agent during their training route and with the same policy update frequency as our shared learning agents. This makes for a more fair comparison and significant improvements to IMPALA as we show later. We label the results for the fleets using these comparison frameworks as 'IMPALA' and 'modified IMPALA'. 

\begin{figure}[h]
     \centering
     \begin{subfigure}{0.44\textwidth}
         \centering
         \includegraphics[width=.75\textwidth]{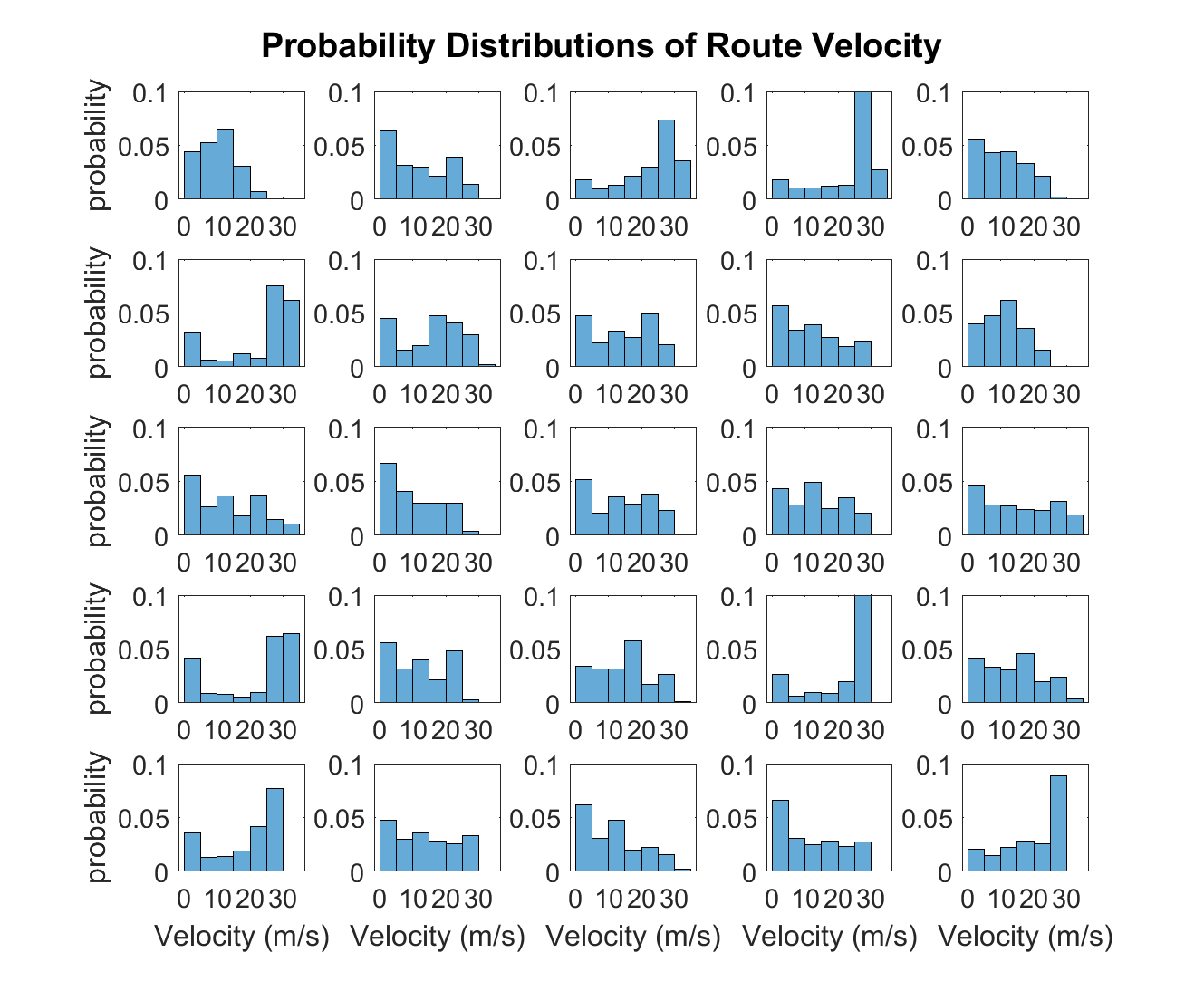}
         \caption{A set of routes ($R$) from real-world data}
         \label{fig:scenario1a}
     \end{subfigure}
	\begin{subfigure}{0.44\textwidth}
		\centering
		\includegraphics[width=.85\textwidth]{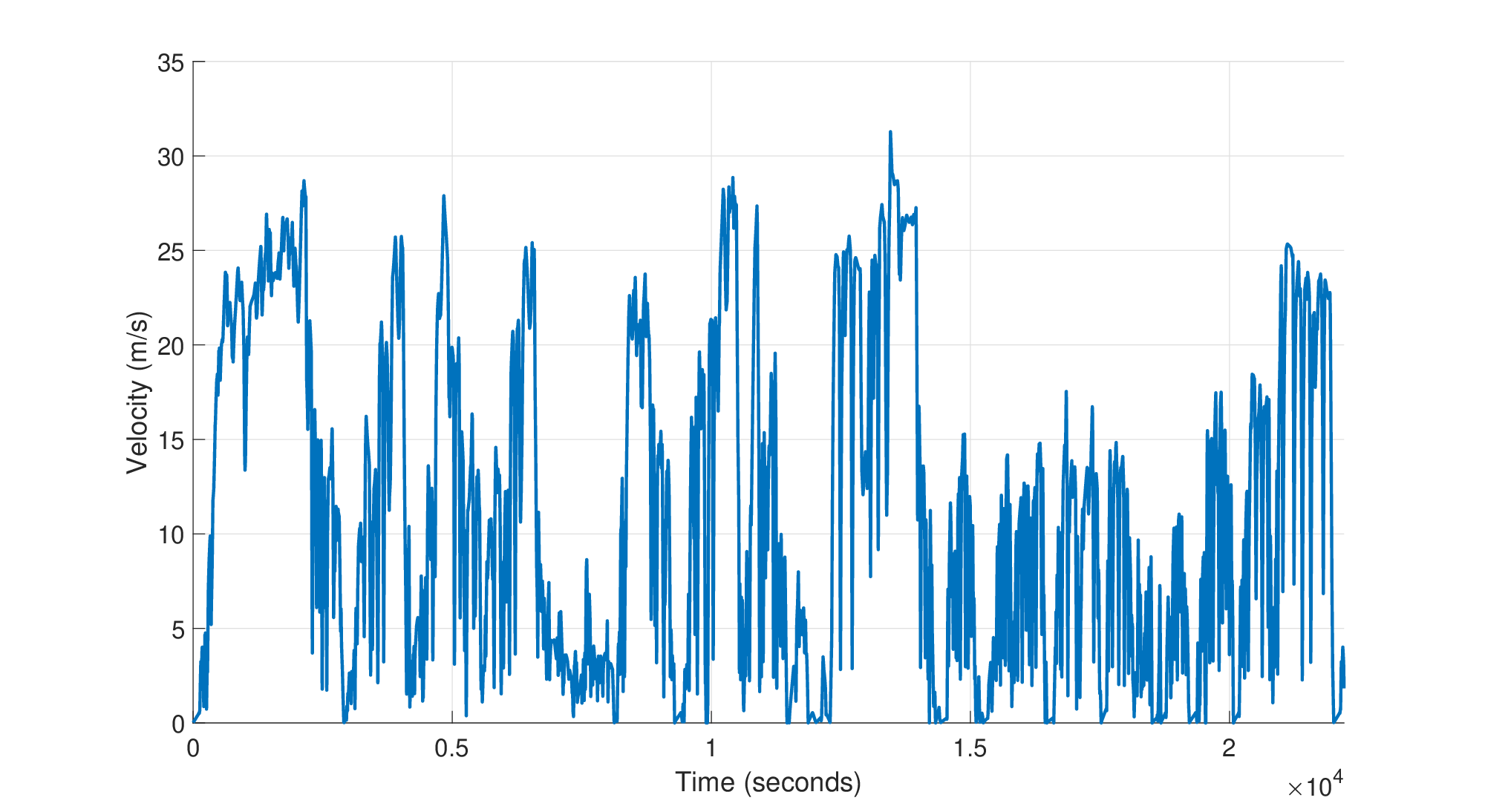}
	      \caption{Velocity profile for the evaluation route}
	      \label{fig:scenario1b}
 	\end{subfigure}
 	\caption{Scenario 1 setup for randomized real-world routes}
	\label{fig:scenario1}	
\end{figure}

\textbf{Scenario I.} In the first scenario, we consider the set of speed profiles ($R$) as 25 drive cycles (routes) randomly selected from telematics data collected from real vehicles. The probability distributions of the velocity profiles characterizing these routes are depicted in Figure \ref{fig:scenario1}, highlighting the diversity within the set of routes. After each training route is completed, the vehicle control policies are evaluated using the greedy (highest-valued) action on a deterministic representative route for the set. This evaluation route merges $3$ of the routes with differing probability distributions of velocity designed to create a representative route for the set ($R$). Figure \ref{fig:scenario1b} shows the velocity profile for the evaluation route used for scenario $1$ to simulate a performance comparison between a fleet using the shared learning framework and a fleet without any shared learning as the baseline.

\begin{figure}[h]
     \centering
     \begin{subfigure}{0.45\textwidth}
         \centering
         \includegraphics[width=.9\textwidth]{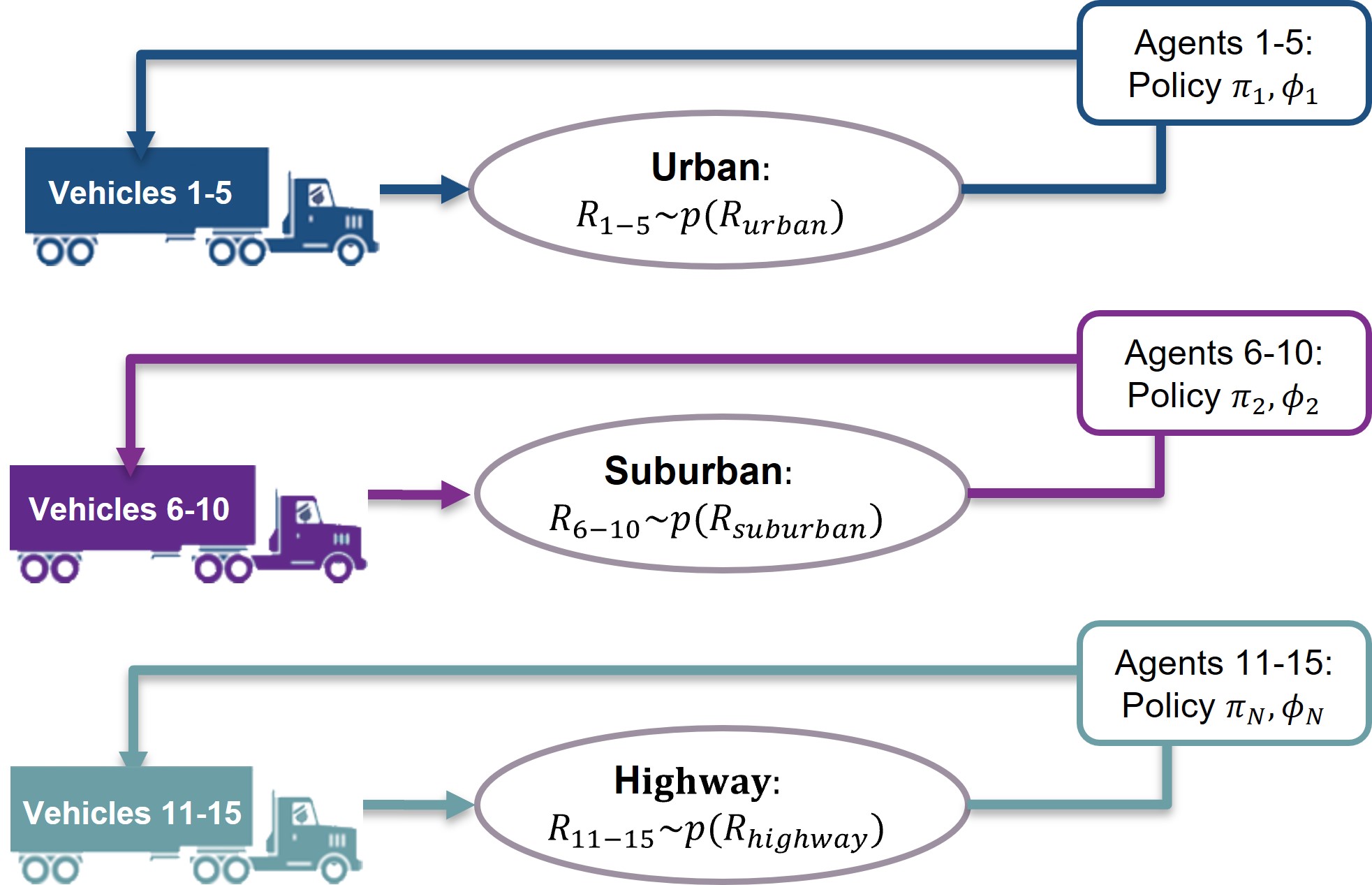}
         \caption{Simulation setup}
         \label{fig:scenario3setupa}
     \end{subfigure}
	\hfill
     \begin{subfigure}{0.45\textwidth}
         \centering
         \includegraphics[width=.99\textwidth]{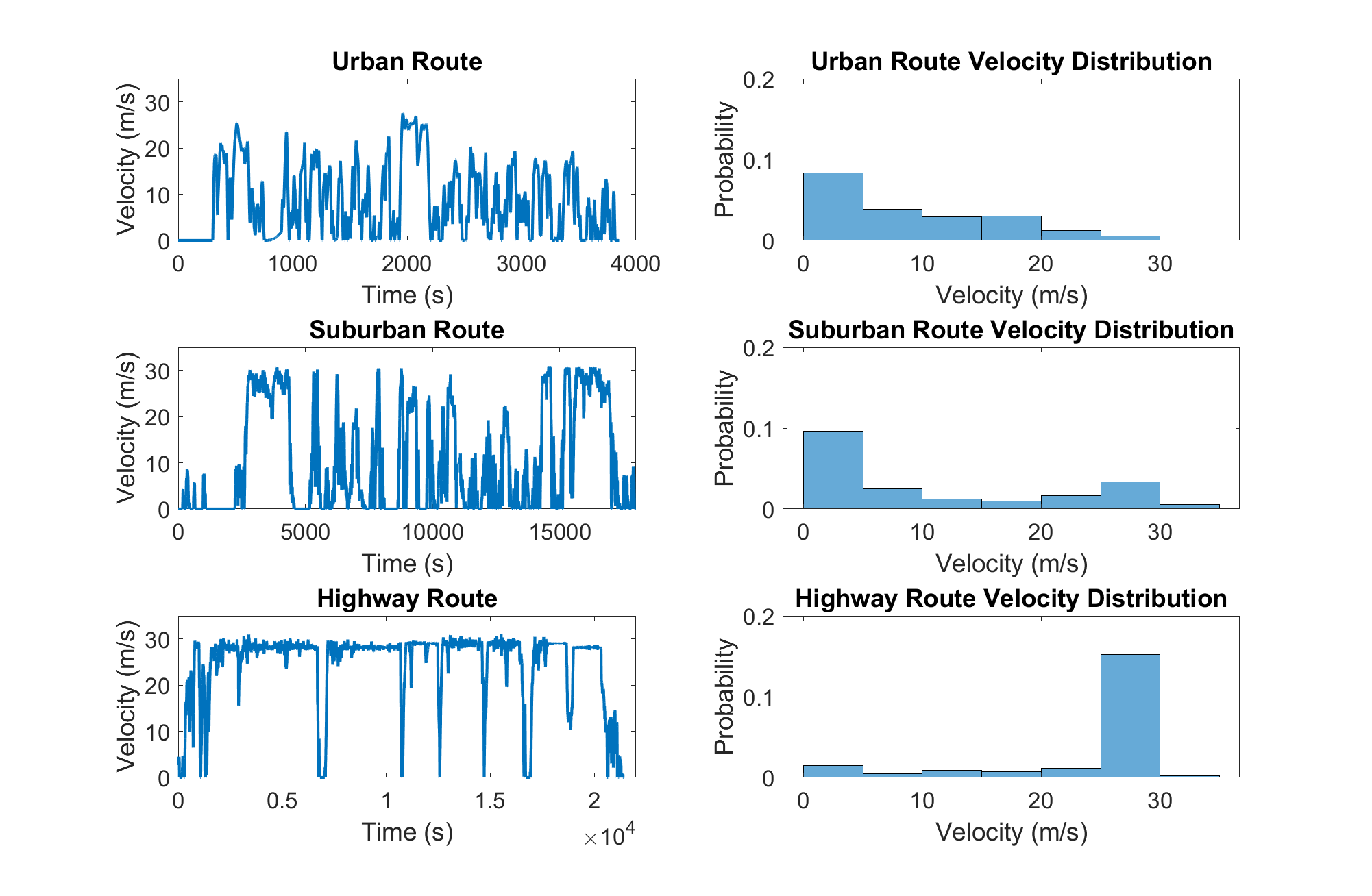}
         \caption{Velocity profiles and distributions}
         \label{fig:scenario3setupb}
     \end{subfigure}
        \caption{Scenario 3 setup for route adaptability (left) and representative routes for $3$ distinct route types (right) }
        \label{fig:scenario3setup}
\end{figure}

\textbf{Scenario II}. In the next scenario, our focus shifts to evaluating the scalability of the shared learning framework with a centralized fleet coordinator, as there are potential computation and data exchange challenges with using a central entity. For this case, we create a random distribution of routes based on the representative suburban route from the DriveCat database~\citep{drivecycles}, shown in Figure \ref{fig:scenario3setupb}. We generate $50$ routes by clustering the velocity and acceleration characteristics of the representative route and randomly sample from these clusters to obtain statistically similar routes~\citep{cycle_syn}. For each training cycle, the vehicles are randomly assigned a route (velocity profile) from the generated set of suburban routes ($R_{suburban}$), while still varying both the IDM driver parameters and vehicle mass. Multiple fleets are simulated across a range of group sizes, while maintaining uniformity in the speed distributions of the routes (specifically suburban routes). Following each training route, each vehicle is evaluated on the representative suburban route to compare performance and computational requirements relative to fleet (group) size.

\textbf{Scenario III}. Lastly, we analyze the agents' ability to adapt to new routes via the shared learning scheme. In the final scenario, we consider a fleet of $15$ vehicles, with $5$ being trained on only urban routes, $5$ on suburban routes, and $5$ on highway routes. Figure \ref{fig:scenario3setup} depicts the fleet setup and the representative routes (taken from the DriveCat database)~\citep{drivecycles} along with their speed distributions used to synthesize a set of $50$ speed profiles for each route type. Similar to our previous setup, each vehicle randomly selects a route from their respective set of speed profiles. The key difference here is that each vehicle's DRL PTC agent is being evaluated on a composition of all 3 distinct route types while exclusively training on their designated set of routes. Subsequently, each vehicle undergoes an evaluation on a route that combines the 3 representative routes resulting in a $45\%$ suburban, $10\%$ urban, and $45\%$ highway route. 

\section{Results and discussion}\label{sec:results}
The presentation of the results are organized following the three scenarios described in the previous section.

\begin{figure}[h]
     \centering
	\begin{subfigure}{0.32\textwidth}
		\centering
		\includegraphics[width=\textwidth]{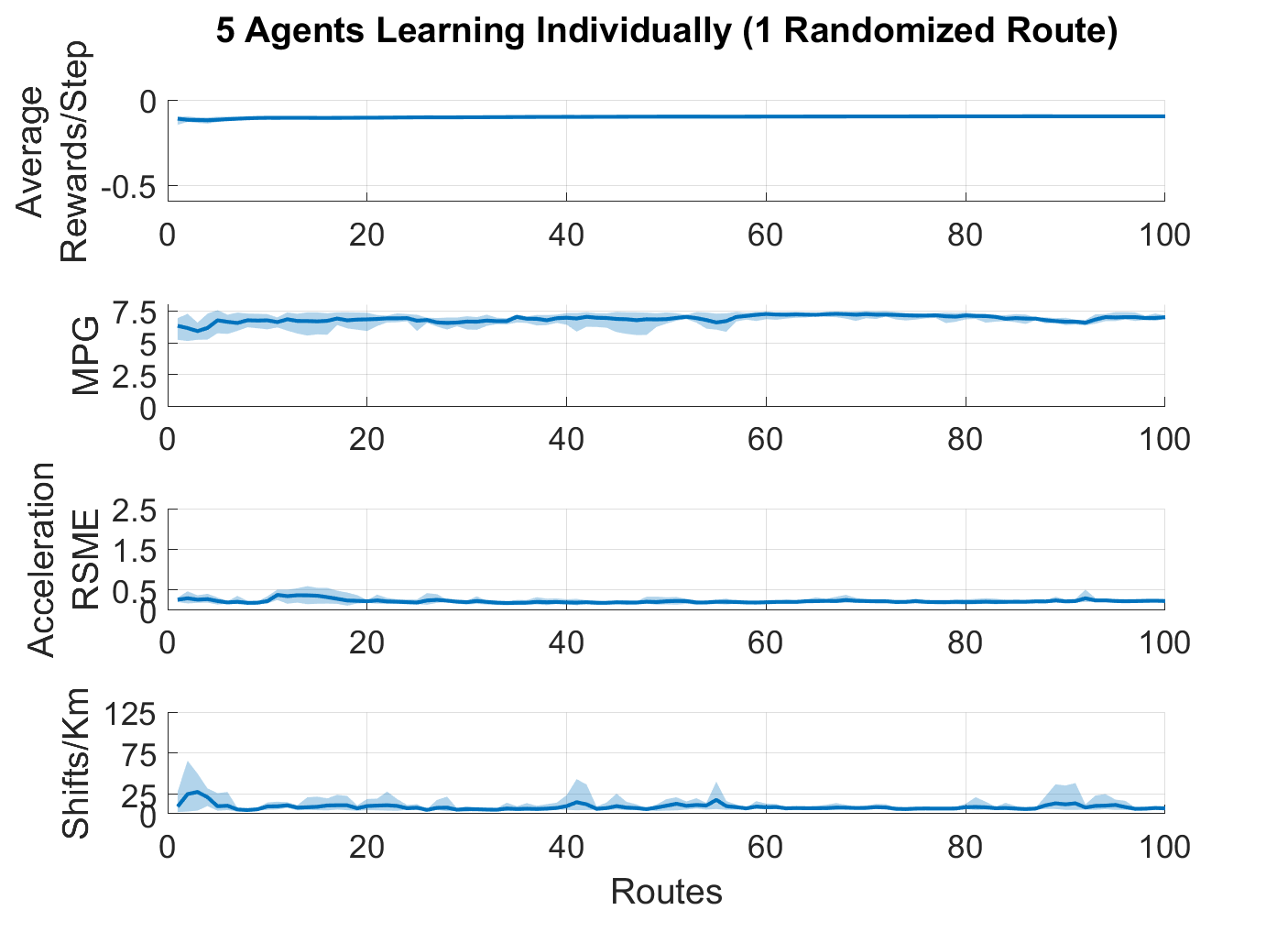}
	      \caption{Individual learning on $1$ route}
	      \label{fig:scenario1resultsa}
 	\end{subfigure}
     \begin{subfigure}{0.32\textwidth}
         \centering
         \includegraphics[width=\textwidth]{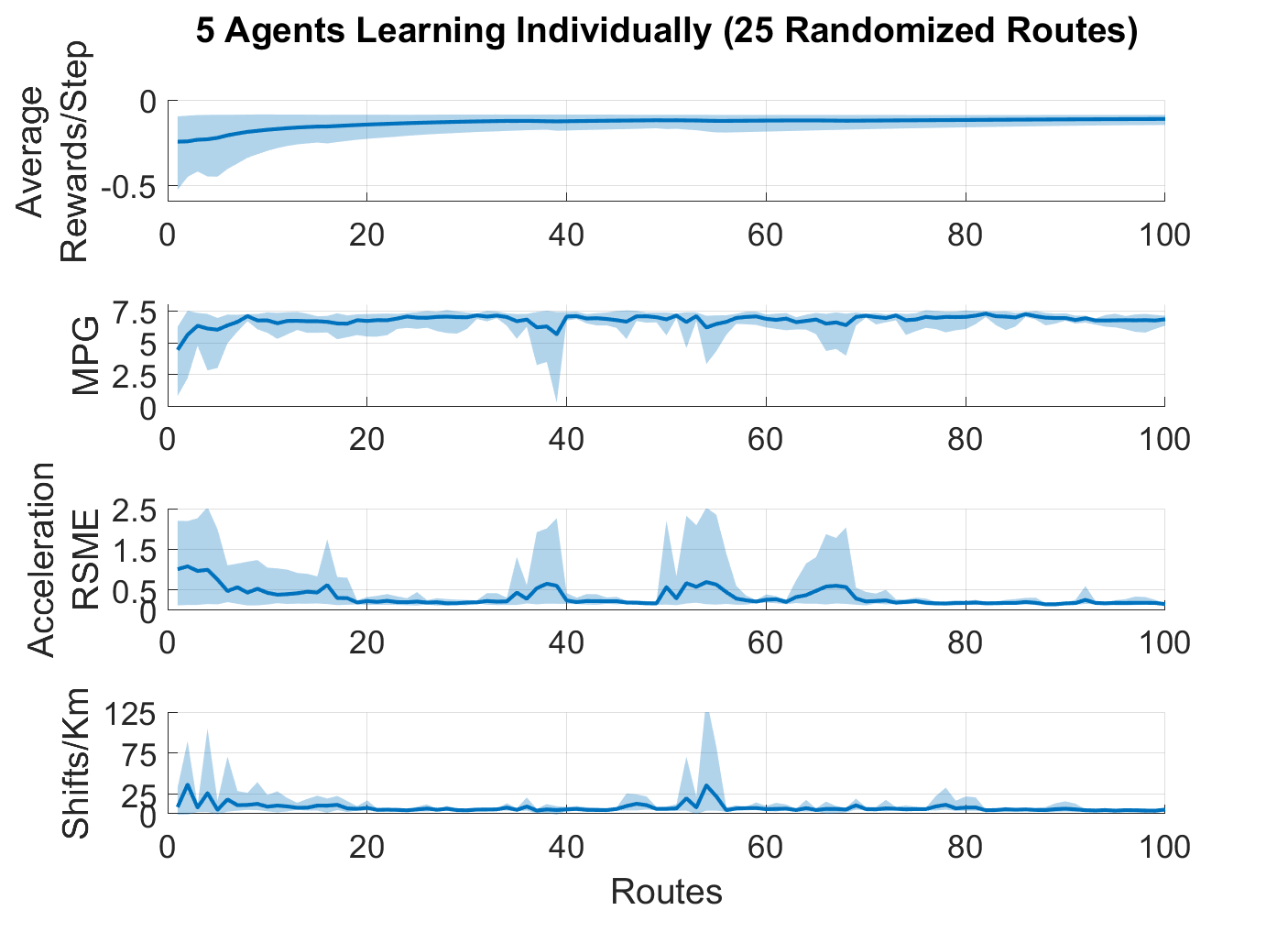}
         \caption{Individual learning on $25$ routes}
         \label{fig:scenario1resultsb}
     \end{subfigure}
     \begin{subfigure}{0.32\textwidth}
         \centering
         \includegraphics[width=\textwidth]{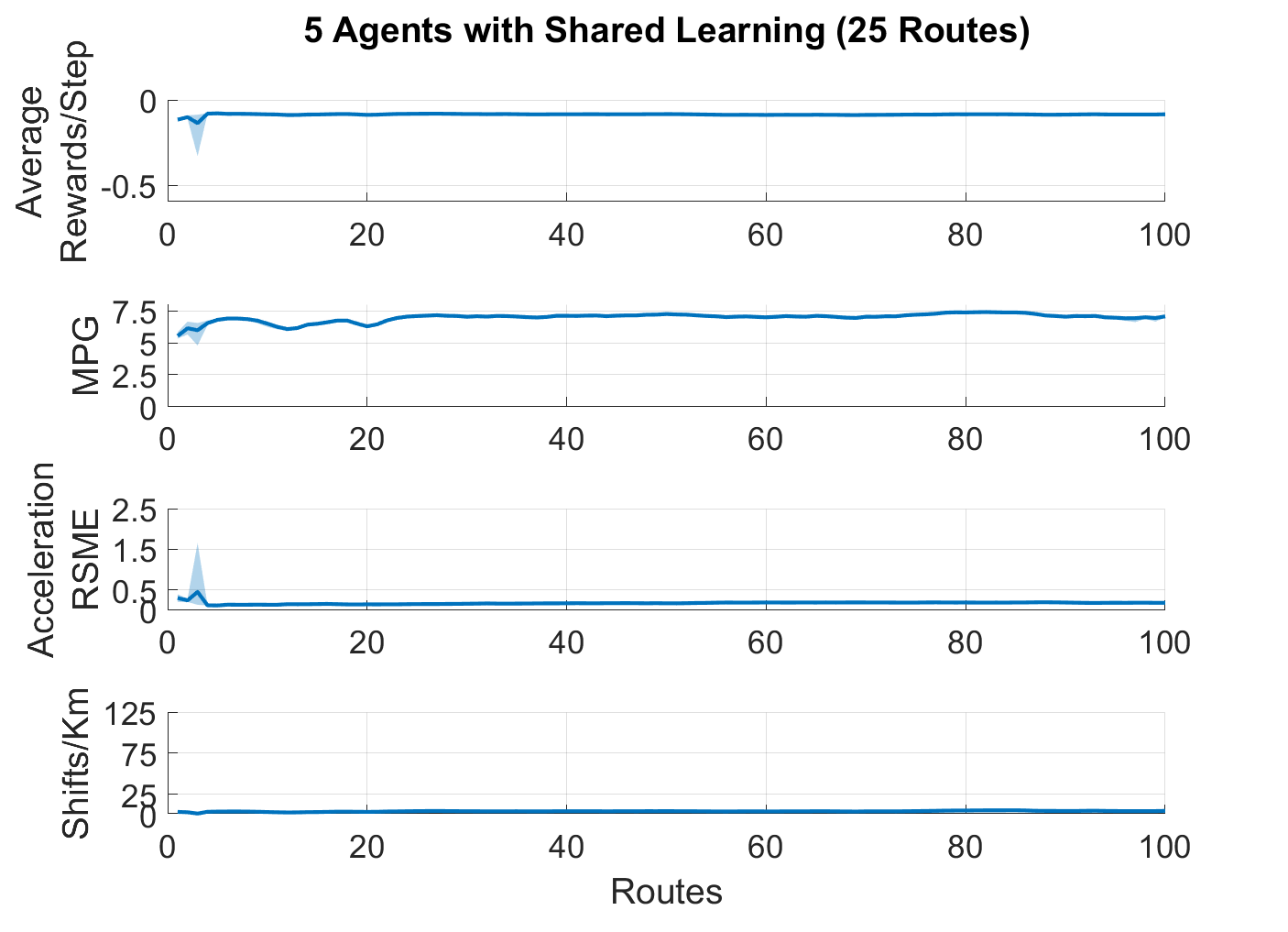}
         \caption{Shared policy learning on $25$ routes }
         \label{fig:scenario1resultsc}
     \end{subfigure}
\par\bigskip
	\begin{subfigure}{0.32\textwidth}
		\centering
		\includegraphics[width=\textwidth]{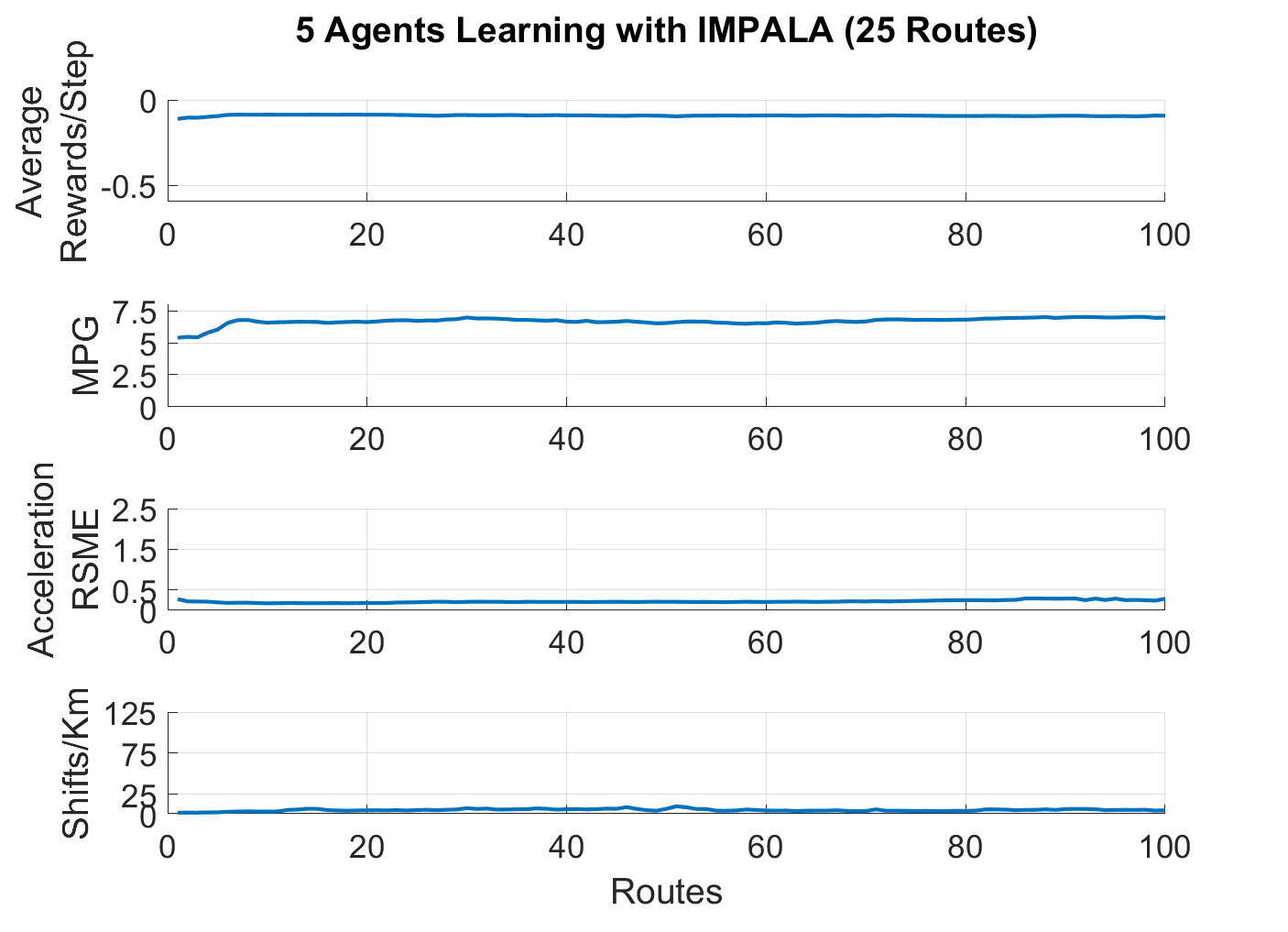}
	      \caption{Learning with IMPALA on $25$ routes}
	      \label{fig:scenario1resultsd}
 	\end{subfigure}
     \begin{subfigure}{0.32\textwidth}
         \centering
         \includegraphics[width=\textwidth]{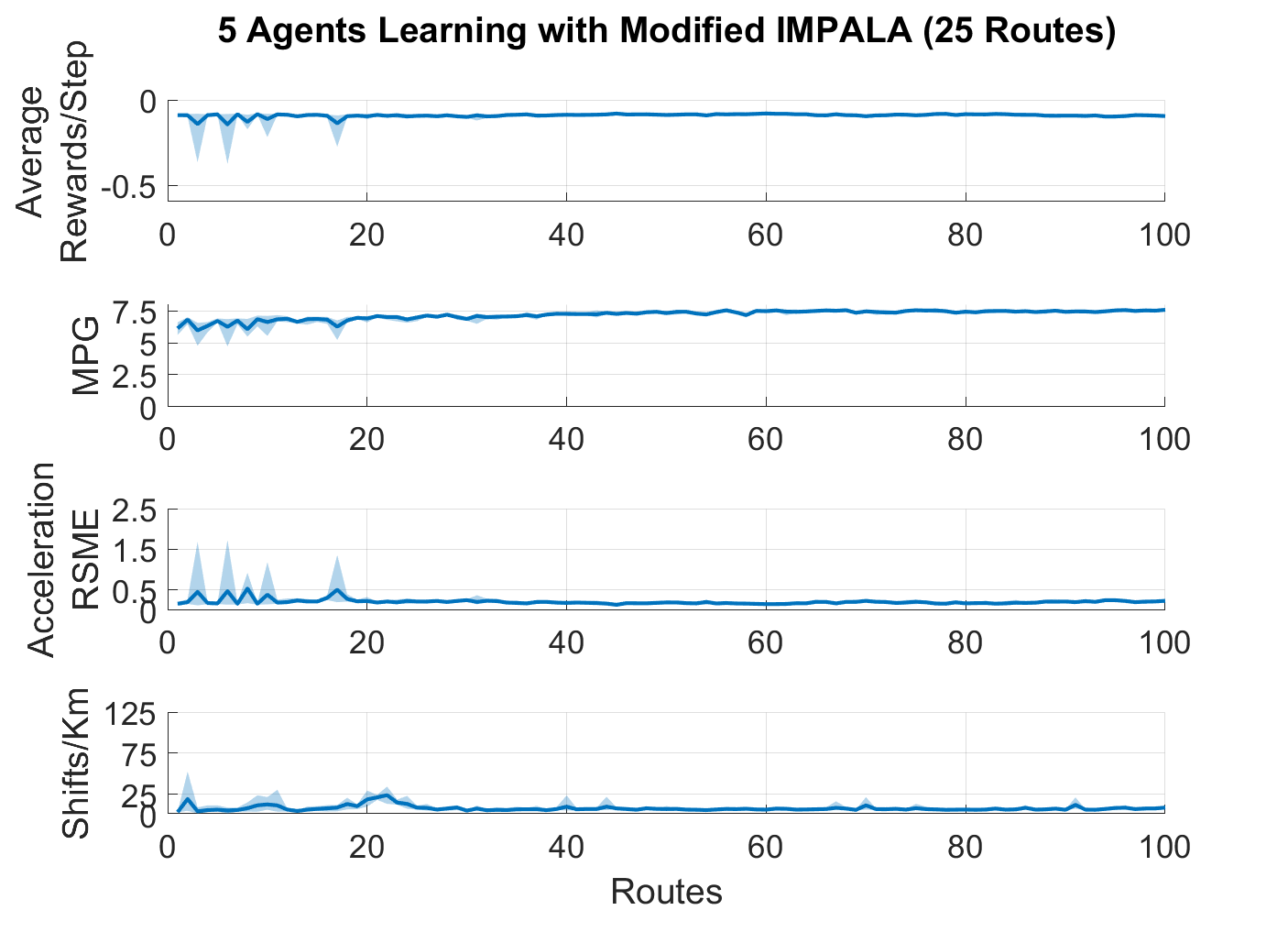}
         \caption{Learning with Modified IMPALA on $25$ routes}
         \label{fig:scenario1resultse}
     \end{subfigure}
     \begin{subfigure}{0.32\textwidth}
         \centering
         \includegraphics[width=\textwidth]{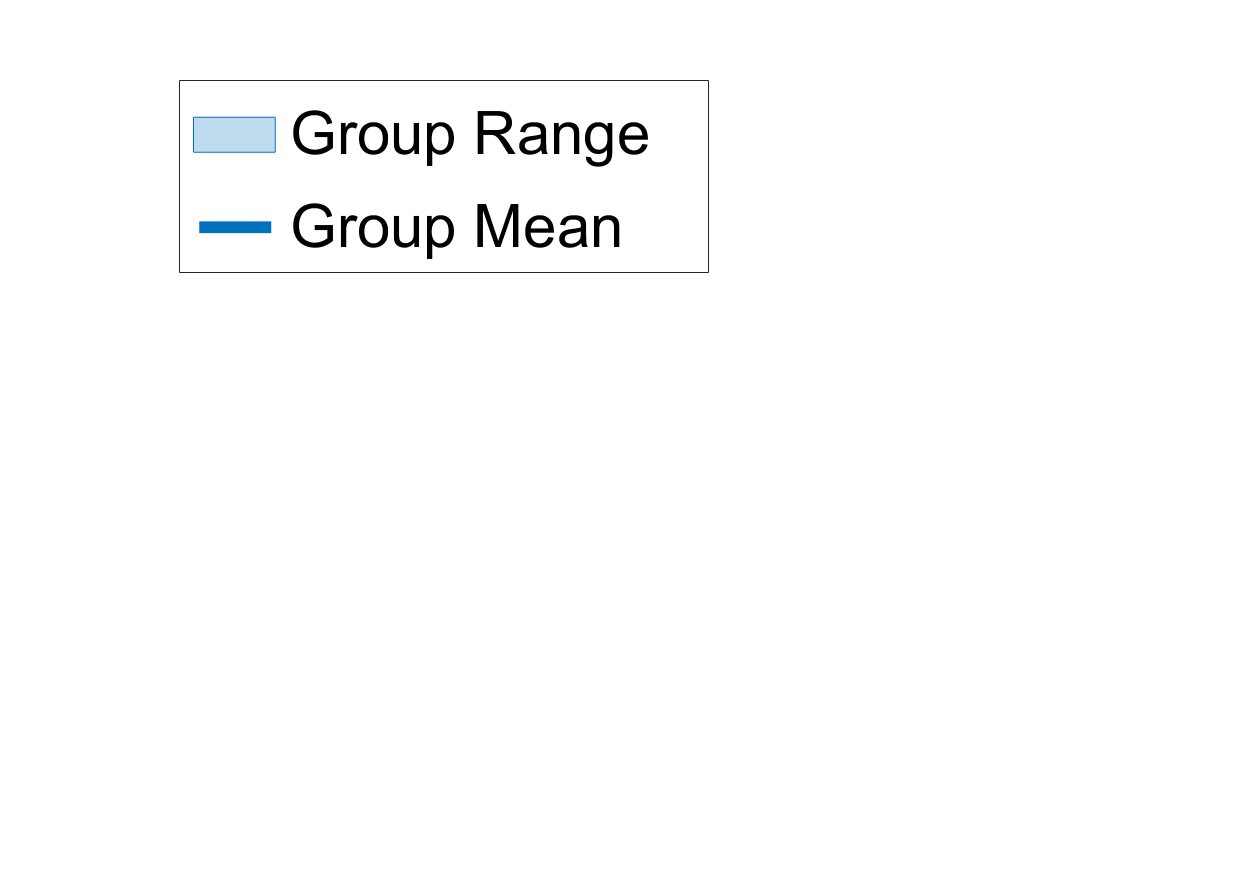}
     \end{subfigure}
        \caption{Shared fleet learning progression using randomized real-world routes comparing fleets of $5$ vehicles}
        \label{fig:scenario1results}
\end{figure}

\textbf{Scenario I.} We first compare the learning progression over 100 routes (cycles) of fleets training on the set of telematics routes. Learned policies are evaluated on the defined evaluation route after each randomly selected training route. The learning curves shown in Figure \ref{fig:scenario1results} represent the mean performance of the fleet, depicted by solid lines, and shaded regions representing the range of the individual agents' performances for each of the evaluations. The performance metrics include the average reward per step, which accounts for all objectives in the reward function in Eq. (\ref{eqn:rewards}) , the fuel consumption (measured in miles per gallon, MPG), the root mean square error (RMSE) between the desired acceleration and the actual vehicle acceleration, and the number of gear shifts per kilometer.  

As a starting point, we include results in Figure \ref{fig:scenario1resultsa} for the common scenario of a DRL PTC agent, where each vehicle repeats the one velocity profile (evaluation route) with added noise to the environment (route, mass, driver) during training and then evaluated with deterministic settings on the same route. Although the initial learning in this setting is stochastic, all agents eventually converge for the deterministic evaluation route with very little variation between agents, meaning the range of the agents' performance is near the group mean as seen in Figure \ref{fig:scenario1resultsa}. This observation doesn't hold when the routes are randomly sampled from the set $R$ as in our baseline fleet scenario (Figure \ref{fig:formulation} and Figure \ref{fig:scenario1a}). With a more diverse set of routes, the individually learning agents show significantly more variation in their performance as demonstrated in Figure \ref{fig:scenario1resultsb} where the range of performance is farther from the mean.   

For the same case of randomly sampled routes with shared learning (Figure \ref{fig:central_framework}), utilizing the group policy significantly reduces the variance in the agents' performance, as shown in Figure \ref{fig:scenario1resultsc}. This can be attributed to the group policy's role in guiding all agents toward adopting common high advantage behaviors, consequently leading to a more uniform performance across the fleet (near the group mean performance).  Figures \ref{fig:scenario1resultsd} and \ref{fig:scenario1resultse} illustrate the alternative shared learning methods of IMPALA and modified IMPALA as described in Section \ref{sec:routingscenarios}. We observe that the original IMPALA has minimal improvement but no variance, as each agent directly uses the updated group policy after each route, while the modified IMPALA shows improved performance overtime as a result of the additional localized learning by each agent. Still, the shared policy learning in Figure \ref{fig:scenario1resultsc} shows the performance with minimal variance. It is worth noting that learning stabilizes with considerably fewer cycles with all shared learning methods compared to the scenario where agents learn independently on the randomly selected routes. 

Figure \ref{fig:scenario1results2} found in \ref{app:results} additionally illustrates the performance of the individual agents of the fleets shown above. It is noticeable that without shared learning, one or two agents often produce poor rewards, showing an unacceptable PTC policy (with large acceleration errors and gear shifting frequency). This does not occur for the fleet using shared learning after 5 routes of training and all agents maintain an acceptable performance. We empirically found that increasing the memory buffer to over $1$ million samples reduces this variance by enabling the five agents to learn on a larger distribution of previous experiences and thus reduces forgetting of prior knowledge from routes previously sampled. This is typically not a viable solution as the real world may have large variations in experiences (more than just velocity profiles considered here) requiring an extremely large buffer leading to slow or even poor learning performance~\citep{memory}.

\begin{figure}[h]
         \centering
         \includegraphics[width=.78\textwidth]{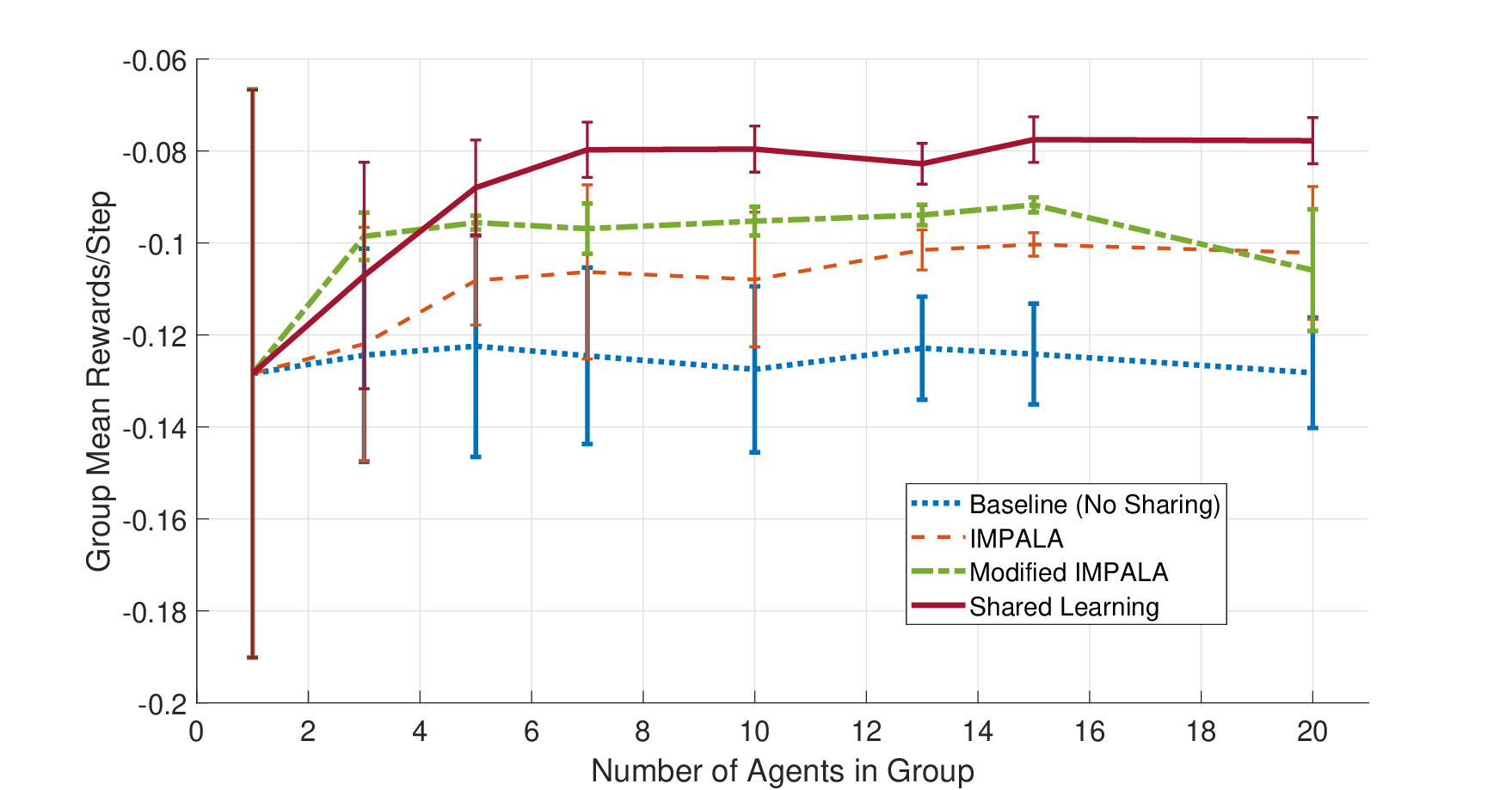}
         \caption{Comparison of overall fleet rewards under three learning schemes relative to the size of the group/fleet}
         \label{fig:scenario2results}
\end{figure}

\textbf{Scenario II}. In the next scenario, we focus on the scalability aspects of shared learning. We evaluate multiple sets of fleets with $1-20$ vehicles for baseline fleets (no sharing), fleets employing IMPALA and the modified version of IMPALA, as described above, and fleets using the proposed shared learning framework. Figure \ref{fig:scenario2results} compares the summary performance of these fleets as we increase the number of vehicles in the group/fleet. For our particular reward function (see Eq. \ref{eqn:rewards}), the rewards are negative (so, the smaller their magnitudes the better). We compute the mean rewards per step across the group of vehicles on an evaluation cycle following each training cycle (sampled from $R_{suburban}$) and then take the average over 100 cycles of training. We fit a normal distribution to the results of multiple group training runs conducted for each group size setting. Figure \ref{fig:scenario2results} shows the corresponding mean and standard deviation of these values as the group size is increased. The size of the sample population for each group size was chosen to maintain a statistical error tolerance below $10\%$ with a $95\%$ confidence level, and below $5\%$ for groups with more than 5 vehicles~\citep{stats}. 

The results in Figure \ref{fig:scenario2results} show that the fleets without shared learning (baseline case with individual learning) maintain a similar mean performance, regardless of the size of the group, as should be expected. The variation is slightly reduced with more agents due to a larger population size used for computing the group mean rewards. The fleets using IMPALA, without localized learning, show only a slight improvement in performance compared to the baseline, particularly with smaller fleet sizes. The modified IMPALA setup, where the agents are allowed local learning updates on their routes, show improved performance over the baseline, but slightly below our proposed framework. The performance of the fleets with modified IMPALA also diminishes as the number of sharing vehicles increases beyond 15, suggesting potential challenges in learning a stable policy from the excessively varied data of all the vehicles in the fleet. For the fleets applying shared learning, the mean reward per step is consistently better than the baseline. With $5$ or more agents, shared learning maintains a better performance than the modified IMPALA. We attribute this improved performance to the advantage weighting of the policies when regressing for the group policy in our framework, rather than learning equally from all data as done with the centralized learner of IMPALA. 

When we consider that the figure shows the group mean reward per step, we note that the gap between the baseline and group learning grows fairly linearly with group size. Furthermore, the variance is reduced with shared learning. This stems from the fact that the individual agents constrain their learned policy to be close to the group policy which suppresses the effect of poorly performing agents through the advantaged regression. We note, however, that the variance in the mean reward doesn't seem to be affected by group sizes larger than 7 or so, indicating that it may not be necessary to have very large groups to take advantage of the shared learning scheme for the suburban scenario simulated. Although this specific threshold may not hold for more diverse route distributions, the observation is noteworthy considering the computational aspects we discuss next.

We evaluate the computational components of the algorithm to understand how the overall algorithm scales with larger group sizes with respect to the computational burden. The algorithmic functions for updating the actor-critic neural networks at individual agents remain unaffected by the number of agents within the group. The only additional function introduced to the base DRL algorithm loss function is the calculation of the KL divergence between the agent's policy and the group policy. We measured the computation time of the individual agent learning algorithm on an NVIDIA RTX A1000 GPU. The mean computation time (over 25 samples) of the learning algorithm was $7.35 \pm 0.8$ seconds without the group policy and $8.03 \pm 0.8$ with the KL loss to the group policy. These times do not change with the number of agents in the group, as is evident from the loss functions in Eq. (\ref{eqn:agent}). If the agent and group policy's network parameter dimensions are kept the same, as we do here, the operations remain identical across all agents, irrespective of fleet size.
\begin{figure}[h]
     \centering
     \begin{subfigure}[h]{0.49\textwidth}
         \centering
         \includegraphics[width=\textwidth]{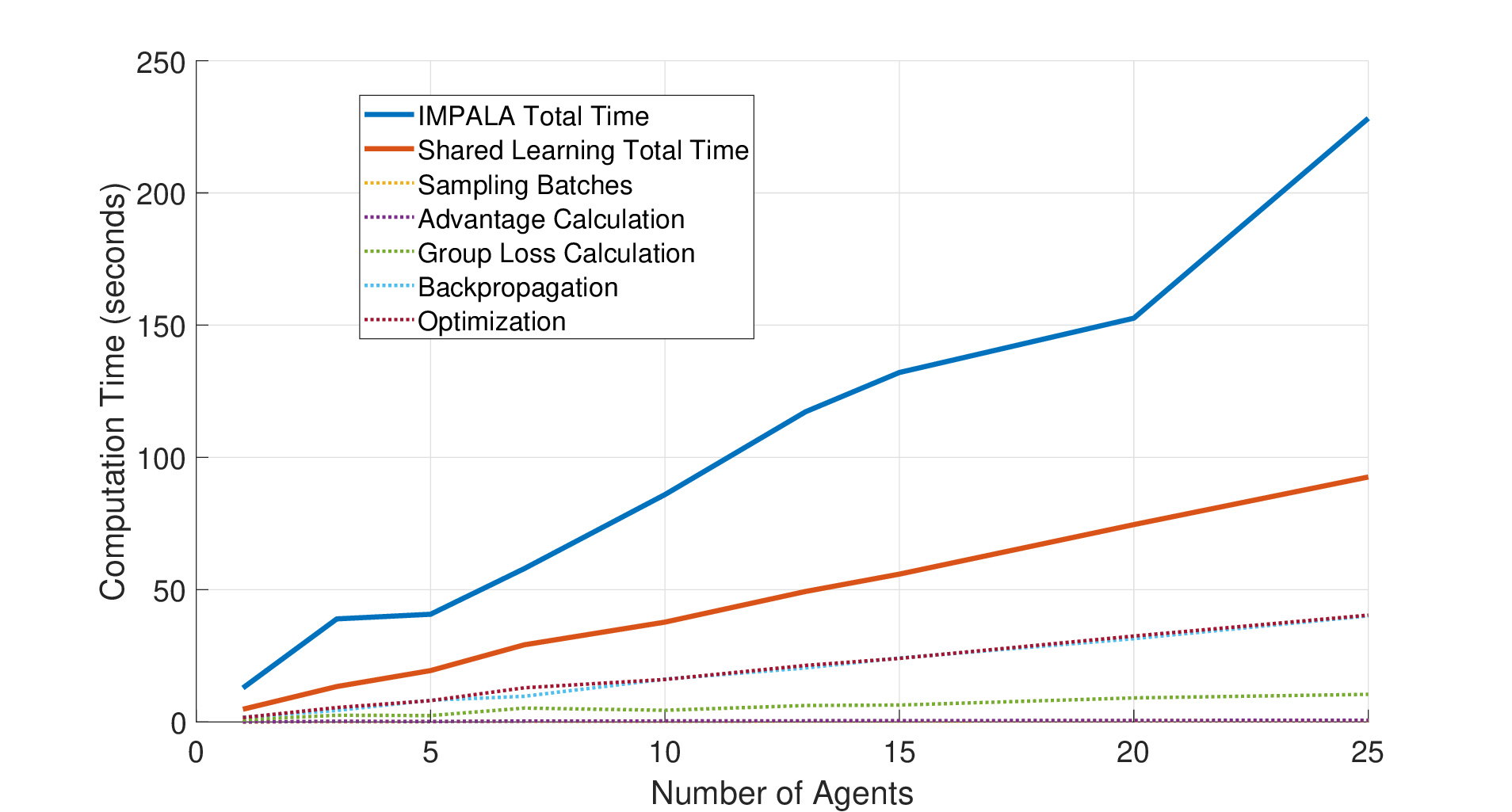}
         \caption{Computation time versus group size}
         \label{fig:comptime}
     \end{subfigure}
     \begin{subfigure}[h]{0.49\textwidth}
         \centering
         \includegraphics[width=.8\textwidth]{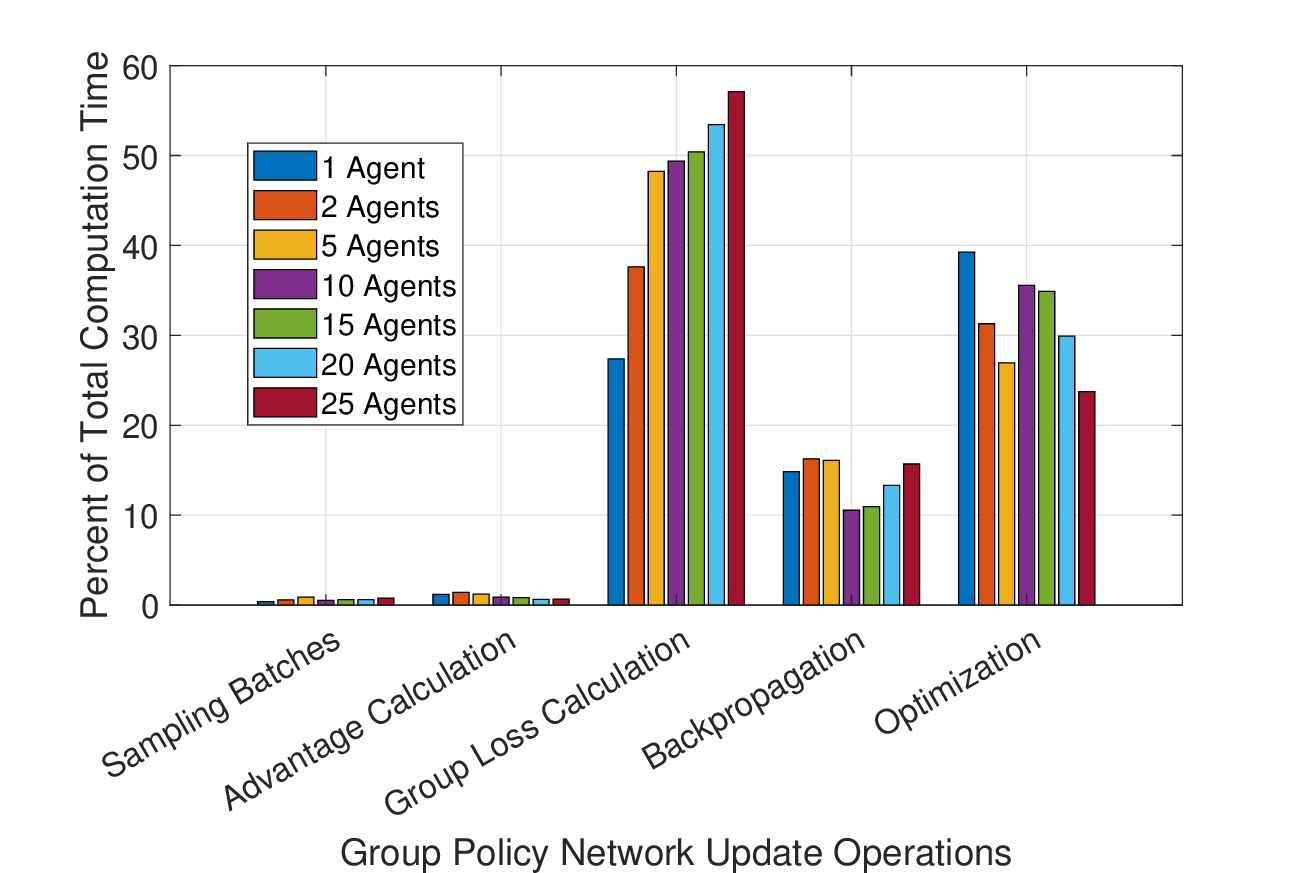}
         \caption{Breakdown by computational components}
         \label{fig:comppercent}
     \end{subfigure}
        \caption{Operations and computation time of updating the group policy network relative to group size}
        \label{fig:scenario2computation}
\end{figure}

The primary potential bottleneck comes from the update of the group policy, as the loss function involves all agent policies, and evidently is a function of the number of agents. To better understand the computational impact of introducing more agents into the group, we deconstruct the group policy regression process into its key components: including batch sampling, computation of the advantage weights, calculating the group loss functions (specifically, the total cross-entropy loss), network backpropagation, and optimization step via the gradient-based Adam optimizer. Figure \ref{fig:comptime} depicts the mean computation time (over 25 computation cycles on the same NVIDIA RTX A1000 GPU) of the central/group policy for IMPALA and shared learning, as well as the computational components of the shared learning for each group size. Modified IMPALA uses the same centralized policy computations as IMPALA. The results in Figure \ref{fig:comptime} show that the total processing time is significantly lower for the shared learning scheme as compared to IMPALA. This should not be surprising as the shared scheme involves policy regression, while IMPALA executes a centralized actor-critic learner. Still, we observe that computation time for shared learning escalates by approximately $25\%$ per each agent added. Figure \ref{fig:comppercent} presents the breakdown of the percentage of the total computational time we observed by implementing the group policy learning regression and measuring on the same device. The computation of the agent loss function has the highest percentage of the total time and increases linearly with the additional cross entropy term per agent. Despite being called only once per update, backpropagation and optimization also show a slight increase in computation time with the addition of more agents.

Data transfer is another potential bottleneck in the centralized shared learning framework. The data transfer from the group policy to the agent policy consists of the updated group policy's ($\pi_g$) neural network parameters, which is independent of the number of agents in the group. To update the group policy, each individual agent must transfer their updated current policy ($\pi_i$) and Q-value ($Q_i$) neural networks along with $B$ states ($\bar{s}_i$) from their corresponding memory buffer. The amount of data transferred to the group policy server (fleet coordinator) increases linearly with each additional agent added depending on the network sizes and the number of states sampled ($B$) per agent. 

To summarize the findings for scalability, although more vehicles in the group may improve the cumulative group performance, the result in Figures \ref{fig:scenario2results} and \ref{fig:scenario2computation} suggests that there are diminishing returns in the mean group reward per step beyond 7 or so vehicles. While this threshold will depend on the task scenario (e.g. diversity of drive cycles sampled), the high computational cost does not seem to justify using larger group sizes for shared learning. However, if computational cost is not a concern, adding more vehicles to the group offers the advantage of keeping more agents closely aligned with the group policy, thereby improving the group total cumulative rewards (the mean rewards from Figure \ref{fig:scenario2results} multiplied by the number of vehicles in the group). This is a linearly growing improvement in performance of the fleet with shared learning over the baseline of individually learning agents.

\begin{figure}[h]
         \centering
         \includegraphics[width=.9\textwidth]{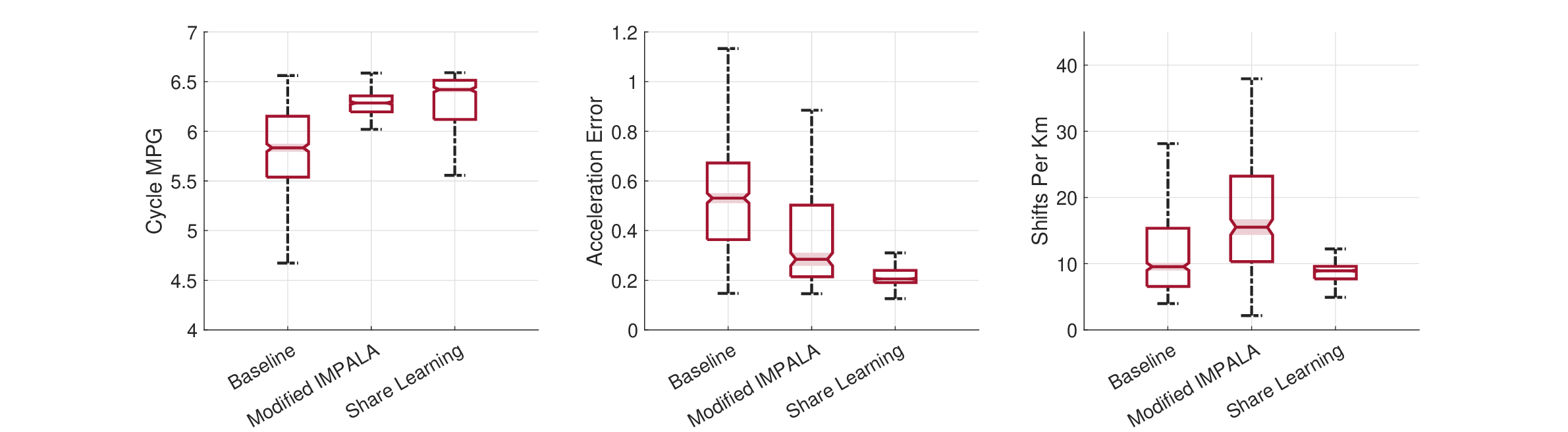}
         \caption{Comparison of the fleet performance parameters evaluated for the 100 routes}
         \label{fig:scenario2performance}
\end{figure}

Using the same dataset for a group size of $10$ vehicles, we next compare the PTC performance distributions among the baseline, the modified IMPALA, and the shared learning framework. In Figure \ref{fig:scenario2performance}, statistics for the cycle MPG, acceleration RMSE, and shifts per kilometer are illustrated for the cumulative evaluations conducted after 100 routes. The mean cycle MPG for the shared learning framework is $8.5\%$ higher on average compared to the fleet of $10$ individually learning agents. Additionally, the performance for the shared learning scheme exhibits a significantly lower acceleration error (mean and variation) compared to the baseline, which is a crucial factor in driver accommodation. One could make similar observations on the metrics of the shifts per km. The modified IMPALA performs some where in the middle with respect to MPG and acceleration error but performs poorly on shift frequency. It is evident that the performance stability for both shared learning frameworks (modified IMPALA, shared learning) is considerably better than the individually learning baseline. Note also that it is possible to emphasize different aspects of the reward components via the weight selections in the reward definition (\ref{eqn:rewards}).


\textbf{Scenario III}. Finally, we analyze the agents' ability to adapt to new routes via the shared learning scheme. As illustrated in Figure  \ref{fig:scenario3setup}, each vehicle exclusively trains on a set of distinct route types (urban, suburban, and highway) and is evaluated on a route representative of all $3$ route types.


\begin{figure}[h]
\centering
\includegraphics[width=.9\textwidth]{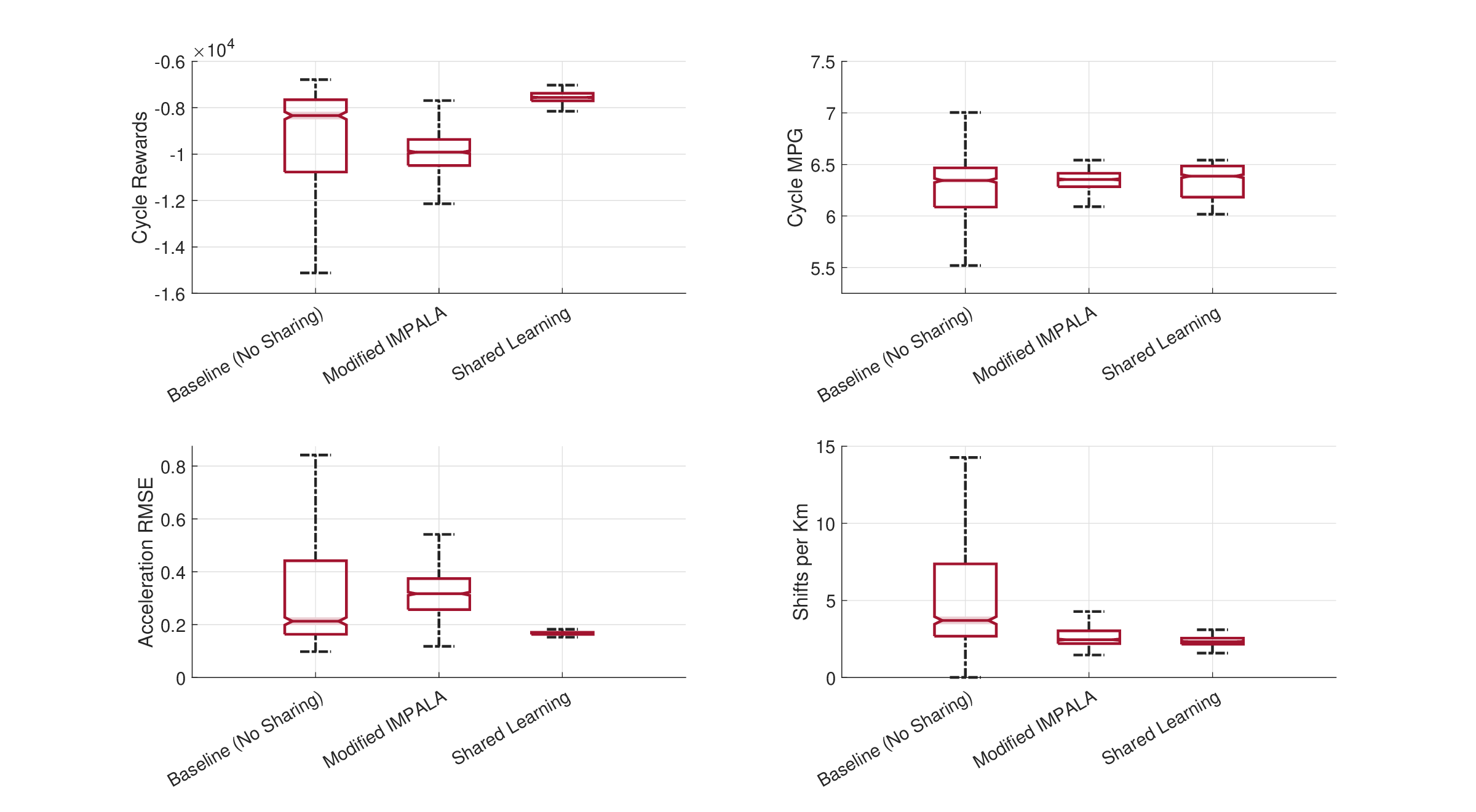}
\caption{Comparison of performance metrics for the adapability scenario}
 \label{fig:scenario3results}
\end{figure}

Figure \ref{fig:scenario3results} demonstrates the performance over $60$ routes for three fleets: one employing individual learning, one using modified IMPALA, and the other utilizing shared learning. The fleet not utilizing a shared learning framework struggles to achieve consistent performance on the evaluation route, shown by an unacceptable range of acceleration error. However, both shared learning methods make it evident that all vehicles are able to learn a consistent performance on untrained route types, even more so when our proposed shared learning framework is employed. The performance, specifically the cycle rewards and the acceleration error, show minimal variation for the fleet. Note that the trade-off between MPG vs. acceleration error vs. shift frequency can be adjusted by selecting different reward weights in the rewards formulation in Eq. (\ref{eqn:rewards}). Still, the observation in Figure \ref{fig:scenario3results} is significant when considering vehicles that are frequently deployed on a wide variety of routes. For example, a vehicle may learn on a specific route for several shifts and then randomly be assigned a completely new route requiring it to quickly adapt. In such cases, the need for policy generalization becomes evident, enabling the agent to operate and adapt to new environments while retaining knowledge from prior experiences of other agents through the group policy.

\section{Conclusions}\label{sec:conclusion}
In this paper, we proposed and demonstrated a shared learning approach for PTC of a fleet of vehicles employing DRL, which is trained from experiences generated in randomly sampled routes serviced by the fleet. The approach distills a group policy from individually acting and learning DRL PTC agents. The group policy is in turn used to regularize the learning of the individual DRL PTC agents. The scheme has the effect of training the DRL PTC agents to learn multi-task policies that seek to maximize performance metrics (fuel economy, driver accommodation/comfort and other metrics) for the distribution of routes that the fleet serves on. The framework can apply for different powertrain technologies and with different state-of-the art DRL algorithms at the individual vehicle level.

The potential benefits of the shared policy learning scheme have been demonstrated by comparing the performance with that of independently learning DRL PTC agents in randomized routes as well as a state-of-the-art approach (IMPALA) that trains a centralized policy using data from all vehicles. It is observed that shared learning improves the mean group reward per step consistently. This translates to a linear growth in the benefits with respect to the group size in terms of cumulative reward for the whole fleet. The variance in the same metric also levels off with a moderate group size of about 7 (although as we noted that can depend on route diversity). These findings are important as it is also observed that the computational times and data traffic grow with group size for the shared learning scheme. Additionally, for a fleet of $10$ vehicles serving a suburban route distribution, the shared learning scheme showed a fleet average improvement of $8.5\%$ in fuel economy over the baseline. This improvement was coupled with a notable reduction in variance across all performance parameters for the fleet, particularly in acceleration error, highlighting the robustness and stability of the performance achieved. Finally, we also demonstrated that the shared fleet learning allows DRL PTC agents to adapt to new environments through experiences shared from other DRL PTC agents.

Future work includes exploring a broader range of tasks beyond speed profiles, such as variations in vehicle models and road grades. Another interesting aspect of the shared framework is experimenting with learning the group policy from a randomly selected subgroup and then applying it to a larger fleet, thereby extending the benefits to more vehicles without the computational burden of learning from all vehicles. More investigations are also needed to ease the computational and data burden for large groups. Lastly, we intend to examine a decentralized sharing platform for a fleet of vehicles to mitigate the computational bottlenecks associated with computing the group policy.

\section*{Acknowledgments}
The authors gratefully acknowledge the support provided by Allison Transmission Inc. for this research.

\bibliographystyle{elsarticle-template-num} 

\bibliography{elsarticle-template-num}

\appendix
\section{DRL algorithm for individual vehicles}
We apply an actor-critic DRL architecture where we first update the critic network which approximates the state-action-value (Q-value) parameterized by $\phi$ for each individual agent.  The Q-value assigns a score to the state and action pair that were taken in the environment using the reward function. The parameters $\phi$ are estimated by minimizing the squared loss function for the error between the network output $Q(s,a|\phi)$ and an estimated target Q-value, $Q_{target}(s_t,a_t)$. We used the retrace $Q_{target}$ from~\cite{ZhuRetrace}), although other targets such as temporal difference (TD)~\citep{sutton} can also be used. 

To update the actor network we apply the MPO algorithm, which uses the base loss function given by:
\begin{equation} \label{eqn:mpobase}
\mathcal{L}_{base} = -\mathbb{E}_{q} \left[ \sum_{t=0}^{\infty} \gamma^t \left[r_t - \alpha KL\left(q(a|s_t)||\pi_{\theta}(a|s_t)\right)\right]\right] - \log{p(\theta)}
\end{equation}
Optimization of the policy is posed as an expectation-maximization problem involving a variational distribution $q(a|s)$.  It is optimized in two steps, where we first optimize for $q$ (in an E-step) by solving for a non-parametric policy that maximizes the critic estimated Q-value while being constrained to remain close to the current policy iterate.

The new policy iterate ($\pi_{\theta}$) is then computed by fitting the parameters $\theta$ to the parametric policy $q(a|s)$ while constraining changes from the current policy.  The objective for this step is written as: 
\begin{equation} \label{eqn:mstep}
\mathcal{L}_{base} =   -\mathbb{E}_{\mu_q(s)}\left[\mathbb{E}_{q(a|s)} \left[\log \pi_{\theta}(a|s) \right) \right] 
\text{ with the constraint }  \; \mathbb{E}_{\mu_q(s)} \left[ KL(\pi_{\theta_k}(a|s)||\pi_{\theta}(a|s))\right] < \epsilon
\end{equation}
where $\mu_{q(s)}$ is the visitation distribution in the buffer, and $\epsilon$ is the bound on the KL contraint. For the hybrid action space of the present application, the KL divergence constraint is factorized into two parts: one for the continous action, and another for the discrete. The reader is referred to ~\cite{neunert2020} for the details of that formulation. 

For updating the agent policies $\pi_i$ in the shared fleet learning setting, where the group policy has been computed as $\pi_g$, we add the forward KL regularization to the loss function as follows:
\begin{equation} \label{eqn:mstep}
\mathcal{L}_{i} = \mathcal{L}_{base} + KL(\pi_g(a|s)||\pi_i(a|s))  \text{ with the constraint }  \mathbb{E}_{\mu_q(s)} \left[ KL(\pi_{\theta_k}(a|s)||\pi_{\theta}(a|s))\right] < \epsilon
\end{equation}
The detailed derivations for the MPO algorithm can be found in~\cite{neunert2020} and~\cite{abdolmaleki2018}.

\section{Simulation Settings}
The simulation and vehicle model parameters are listed on the left in Table 1 and the DRL hyperparameters for individual agent policy learning and the group policy regression in the shared setting are listed on the right.
\captionof{table}{Vehicle, driver model, and DRL settings}
	
\begin{tabular}{ |cc|cc|cc|cc |}
			\hline
			\multicolumn{4}{|c|}{Vehicle Model and Simulation Parameters} & \multicolumn{4}{|c|}{Hyper-parameters for Shared Learning} \\
			\hline\hline
			 Mass & $8000-24000 kg$  & $T_{t,max}$ &  $18 kNm$ & actor learning rate & $1e^{-4}$ &    retrace steps & $6$ \\
			\hline
			$A_f$ & $7.71 m^2$  & IDM $b, \delta$ & $3, 4$ & critic learning rate & $5e^{-4}$ & $\beta$ & $0.8$\\
			\hline
			$r_{wheel}$ & $0.498$  & $t_h$ & $3-4 s$ & group learning rate & $5e^{-4}$ & $\lambda_{d, i}, \lambda_{c, i}$ & $0.6, 0.8$ \\ 
			\hline
			$C_r, C_d$ & $0.015, 0.8$  & $\Delta t$ & $0.5 s$ & Memory Size & $300000$  & $\epsilon_{g_d}, \epsilon_{g_{\mu}}$ & 0.05, 0.05$$\\
			\hline
			$R_{fd}, \eta_{fd}$ & $2.64, 93.7\%$  & $s_0$ & $5-7 m$ &$\epsilon_{\mu}, \epsilon_{\sigma}, \epsilon_{d}$ & $.1, .001, .1$ & $\xi_c, \xi_d$ & $7e^{-2}, 5e^{-2}$\\ 
			\hline
			$\dot{m}_{f,max}$ & $18 g/s$ & $A_{max}$ & $3 m/s^2$ & Regression Iterations & $300$ & batch size & $3072$  \\
			\hline
			&& && $\gamma$, $\lambda$ & $.99$, $1.0$ & $N$ batches  & $20$ \\
\hline
			&&& &$\tau_c, \tau_a$ & $0.01, 0.95$ &$M$ samples & $30$\\
\hline
\end{tabular}
\section{Additional Results}\label{app:results}
Expanding on the results from Figures \ref{fig:scenario1resultsb} and \ref{fig:scenario1resultsc}, we present the performance of the individual agents from the same data. This demonstrates that, typically, only one or two agents at a time exhibit poor performance in the fleet without shared learning. In the shared learning case, all agents perform very similarly to each other.
\begin{figure}[h]
     \centering
     \begin{subfigure}{0.49\textwidth}
         \centering
         \includegraphics[width=\textwidth]{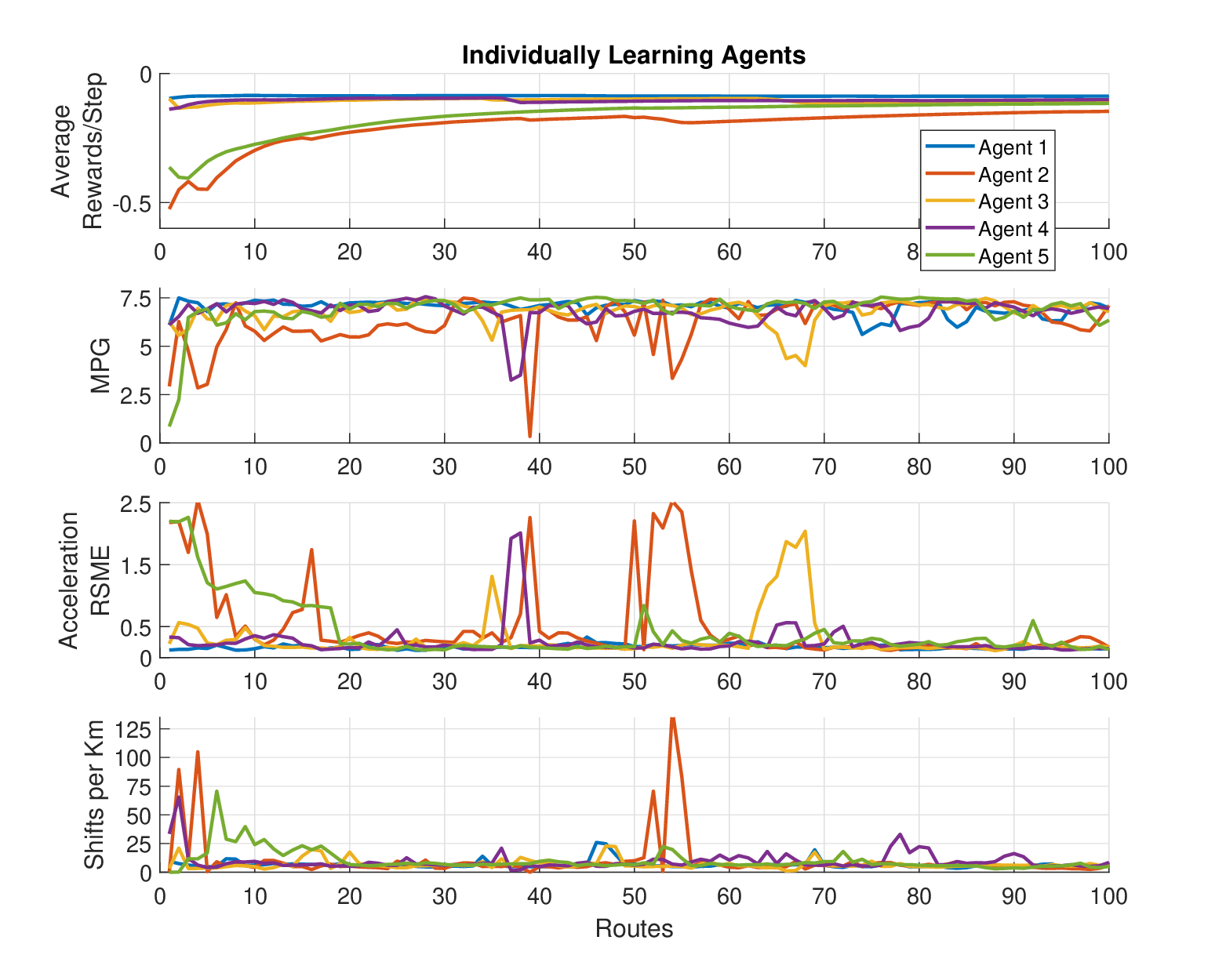}
         \caption{Each agent with individual learning (baseline)}
         \label{fig:s1resultsa}
     \end{subfigure}
     \begin{subfigure}{0.49\textwidth}
         \centering
         \includegraphics[width=\textwidth]{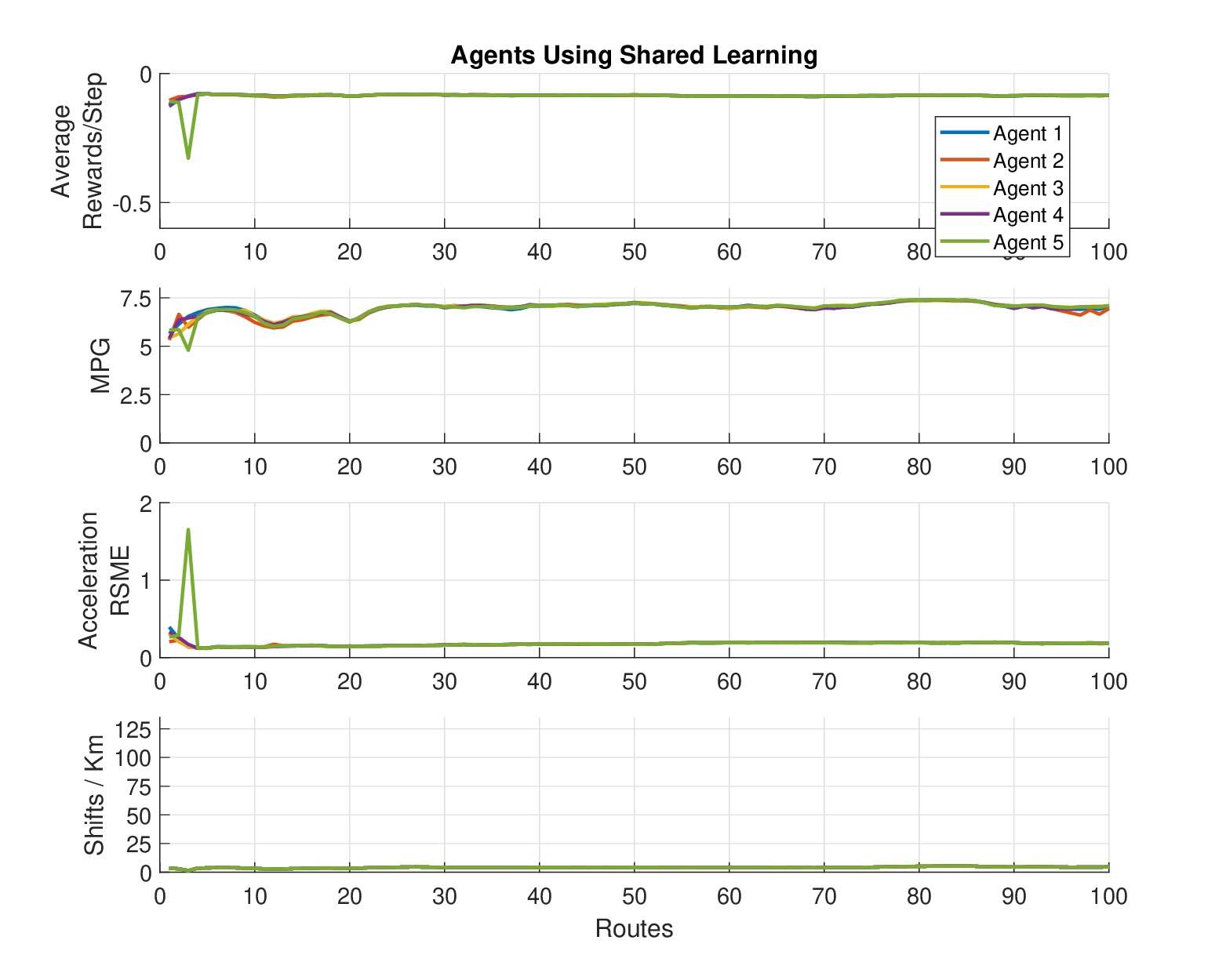}
         \caption{Each agent with shared learning}
         \label{fig:s1resultsb}
     \end{subfigure}
        \caption{Learning progression for individual agent performance using randomized real-world routes}
        \label{fig:scenario1results2}
\end{figure}
\FloatBarrier
\end{document}